\documentclass[useAMS,usenatbib]{mn2e}
\pdfoutput=1
\usepackage{epsfig}
\usepackage{graphicx}
\usepackage{epsf,rotating}
\usepackage{amsmath}
\usepackage{subfigure}
\usepackage{lscape}
\usepackage{comment}
\usepackage[T1]{fontenc}

\usepackage{natbib}

\newcommand{\HI}{H{\sc I}}

\newcommand{\hmpc}{h^{-1}{\rm Mpc}}

\newcommand{\kms}{\;{\rm km}\,{\rm s}^{-1}}
\newcommand\cdunits{{\rm cm}^{-2}}

\newcommand{\gad}{{\sc Gadget-2}}
\newcommand{\autovp}{{\sc AutoVP}}
\newcommand{\ion}[2]{\hbox{#1\,{\sc #2}}}
\newcommand{\vw}{v_{\rm w}}

\newcommand{\apj}{ApJ}
\newcommand{\mnras}{MNRAS}
\newcommand{\apjs}{ApJS}

\newcommand{\msolar}{\;{\rm M}_{\odot} }

\newcommand{\mlo}{\;{${10}^{11}M_\odot$}}

\newcommand{\mhi}{\;{${10}^{12}M_\odot$}}

\title[Inflows and Outflows]{Tracing Inflows and Outflows with Absorption Lines 
in Circumgalactic Gas}  

\author[Ford et al.]{
\parbox[t]{\textwidth}{\vspace{-1cm}
Amanda Brady Ford$^1$, Romeel Dav\'{e}$^{2,3,4,1}$, Benjamin D. 
Oppenheimer$^{5,6}$, Neal Katz$^6$, Juna A. Kollmeier$^7$, Robert Thompson$^1$, David H. Weinberg$^8$}
\\\\$^1$ Astronomy Department, University of Arizona, Tucson, AZ 85721, USA
\\$^2$ University of the Western Cape, Bellville, Cape Town 7535, South Africa
\\$^3$ South African Astronomical Observatories, Observatory, Cape Town 7925, 
South Africa
\\$^4$ African Institute for Mathematical Sciences, Muizenberg, Cape Town 7945, 
South Africa
\\$^5$ Leiden Observatory, Leiden University, PO Box 9513, 2300 RA Leiden, 
Netherlands
\\$^6$ CASA, Department of Astrophysical and Planetary Sciences, University of Colorado, Boulder, CO 80309, USA
\\$^7$ Astronomy Department, University of Massachusetts, Amherst, MA 01003, USA
\\$^8$ Observatories of the Carnegie Institution of Washington, Pasadena, CA 
91101, USA
\\$^9$ Astronomy Department and CCAPP, Ohio State University, Columbus, OH 
43210, USA
}

\begin{document}

\pubyear{2013}
\label{firstpage}

\maketitle
\begin{abstract}
We examine how \HI\ and metal absorption lines within low-redshift
galaxy halos trace the dynamical state of circumgalactic
gas, using cosmological hydrodynamic simulations that include a
well-vetted heuristic model for galactic outflows.  We categorize
inflowing, outflowing, and ambient gas based on its history and
fate as tracked in our simulation.  Following our earlier work
showing that the ionisation level of absorbers was a primary
factor in determining the physical conditions of absorbing gas, we
show here that it is also a governing factor for its dynamical
state.  Low-ionisation metal absorbers (e.g. \ion{Mg}{ii}) tend to
arise in gas that will fall onto galaxies within several Gyr, while
high-ionisation metal absorbers (e.g. \ion{O}{vi}) generally trace
material that was deposited by outflows many Gyr ago.  Inflowing gas
is dominated by enriched material that was previously ejected in
an outflow, hence accretion at low redshifts is typically substantially
enriched.  Recycling wind material is preferentially found closer
to galaxies, and is more dominant in lower-mass halos since high-mass halos have more hot gas that is able to support itself against
infall.  Low-mass halos also tend to re-eject more of their accreted
material, owing to our outflow prescription that employs higher
mass loading factors for lower-mass galaxies.  Typical \HI\ absorbers
trace unenriched ambient material that is not participating in the baryon
cycle, but stronger \HI\ absorbers arise in cool, enriched inflowing gas.
Instantaneous radial velocity measures of absorbers are
generally poor at distinguishing between inflowing and
outflowing gas, except in the case of very recent outflows.
These results suggest that probing halo gas using a range of absorbers
can provide detailed information about the amount and physical
conditions of material that is participating in the baryon cycle.
\end{abstract}

 \section{Intro}

The modern view of galaxy formation relies on continual inflows of
gas from the intergalactic 
medium~\citep[e.g.][]{ker05,bro09,dek09,bou10,dav12,vdv12},
counteracted by strong galactic-scale
outflows~\citep[e.g.][]{opp10,dav11a}, which in concert establish
the growth rate of gas and stars within galaxies at all cosmic
epochs.  This ``baryon cycle" view of galaxy formation has alleviated
many of the classic problems in galaxy formation such as
overcooling~\citep[e.g.][]{whi91,bal01,spr03} and the formation of overly 
compact
disk galaxies~\citep[e.g.][]{gov07,bro12}.  The amount of ejected material
in these types of models is typically
comparable to or exceeds the star formation rate~\citep[e.g.][]{opp08},
and thus the inflow rate is expected to be several to many times
the star formation rate.

Despite the large amounts of mass purportedly moving in and out of
galaxies, direct comparison between observations and theoretical predictions remain difficult.  Outflows are seen emanating from low ~\citep [e.g.][]{mar05, rup05, tre07}, 
intermediate~\citep[e.g.][]{wei09, rub12} and
high~\citep[e.g.][]{pet01,ste01,vei05} redshift star-forming 
galaxies, but the amount of mass involved is difficult to
estimate directly~\citep[e.g.][]{gen10,mar13} owing to complex
ionisation conditions and the tenuous, multi-phase nature of the
outflowing material.  Inflows, meanwhile, are even more difficult
to detect, and there are only tantalising hints from redshifted
IGM absorption lines~\citep{rub12}, though it has been argued
that Lyman-alpha blobs~\citep{goe10}, Lyman limit absorption
systems~\citep[e.g.][]{fum11,leh12}, low metallicity gas~\citep{len13} and high column density \ion{H}{i} 
systems~\citep{vdv12} may be indirect markers of
such accretion, and \citet{fau11} argue that cold accretion streams would have 
very low covering fractions.  Nonetheless, simple arguments from observed 
galaxy properties such as the gas-phase metallicity~\citep[e.g.][]{erb06}
and the evolution of the gas content~\citep{tac10} strongly
suggest that inflows and outflows must be occurring.  Clearly, a
better understanding of the physical processes involved is crucial
for establishing a complete view of how galaxies evolve across
cosmic time. This paper examines the dynamics and origin of accreting and outflowing gas in a cosmological hydrodynamic simulation, at redshift z=0.25, and discusses absorption-line diagnostics that can trace these components. 

A separate aspect of cosmic baryons influenced by galactic outflows is 
the metal content of the intergalactic medium (IGM).  Intergalactic metals
are seen in quasar absorption line spectra from almost the earliest
epochs where such data can be
obtained~\citep[e.g.][]{son01,bec11,sim11}, as well as at lower 
redshift~\citep[e.g.][]{tri00,tri08,tho08,dod10,tum11,coo13,til12,wer13}, 
indicative
of enrichment by strong galactic outflows~\citep{agu01,opp06}.
IGM enrichment observations can thus place tight constraints on the
properties of outflows~\citep{opp06,wie10,opp12}.  However, the
enrichment observed far from galaxies may only track the ancient relics
of outflows, rather than the actively ongoing baryon cycle.
Also, such metal lines may not trace inflows effectively if the
accreting material is metal-poor.

Recently, much attention has been focused on the circumgalactic
medium (CGM), a loosely-defined term that can alternately mean the
gas within some fixed distance of a galaxy~\citep[e.g.][]{rud12,for13},
within the typical metal-enriched region around a
galaxy~\citep[e.g.][]{pro13}, or within the virial radius of
the galaxy's dark matter halo~\citep{sto13}.  The CGM is where the baryon cycle is expected to be
in action, where material is both flowing into galaxies to fuel star formation,
and being expelled from galaxies on its journey into the IGM.
Hence a potentially powerful tool to explore the baryon cycle is
to examine inflowing and outflowing gas in the CGM, either in
emission or, as we discuss here, in absorption against background
sources whose lines of sight pass through a galaxy's CGM.

The recent installation of the Cosmic Origins Spectrograph aboard {\it
Hubble} has enabled the CGM to be probed in absorption in unprecedented
detail.  As an example, the COS-Halos project \citep{tum11,tum13}
examines the CGM around more than 70 galaxies spanning a range
in mass and colour, providing a comprehensive look at the CGM within
$\sim$150~kpc of galaxies ranging in size from $L^*$ to Magellanic-sized 
dwarfs~\citep{wer13}.
Other studies have targeted randomly situated bright quasars but
have obtained extensive galaxy redshift information along their
lines of sight, and are thus able to probe the CGM out to several
hundred kpc~\citep[e.g.][]{tri06,sto13}.  The \ion{Mg}{ii}
doublet, which redshifts into the optical at $z\ga 0.2$, has long
been used as a probe of CGM
gas~\citep{chu00,bor11,che12,kac12} out to many
tens of kpc.  These studies have already highlighted some interesting
results, including the fact that the presence of \ion{O}{vi} is
highly correlated with specific star formation rate~\citep{tum11},
but the presence of low-ionisation lines such as strong \HI~\citep{tho12},
\ion{Mg}{ii}~\citep{che12}, and \ion{Si}{iii}~\citep{wer13} is
not.

One would like to assemble the multitude of observed absorption
features, from \HI\, to low-ionisation IGM metal lines, to high-ionisation
lines that presumably trace more diffuse or hotter gas, into a coherent
picture for how the baryon cycle operates. To properly
place these data within a cosmological context, interpretive models must be
cosmological in nature, and they must explicitly include galactic outflows
and properly account for inflows.  However,
modelling the detailed physics of the interaction between inflowing,
outflowing, and ambient gas remains a great challenge.  Inflows are
thought to be filamentary but some simulations predict
highly collimated and cold filaments~\citep[e.g.][]{dek09}, 
others suggest more diffuse, warmer filaments~\citep[e.g.][]{tor13},
while still others suggest that filaments may break up via thermal
instabilities depending on the environment~\citep{kerhern09}.  Outflows
also have large modelling uncertainties, as the way in which outflows
are powered is still not fundamentally understood, and different
approaches can lead to different results regarding the nature of the
CGM gas and its absorption signatures~\citep{sti12,hum13,for13}.
Finally, it is possible that conductive interfaces on cold clouds
moving through the CGM may be responsible for high-ionisation metal
lines such as \ion{O}{vi} and \ion{Ne}{viii} in
particular~\citep[e.g.][]{tri11}, and such interfaces would be well
beyond the ability for any current cosmological simulation
to resolve.  Hence, these are early days in understanding
how to interpret CGM absorption data, and much work remains.  
Nonetheless, some basic characteristics of the CGM and its relation
to the baryon cycle are likely to be robust in such models.

In this series of papers, we explore the physical conditions and
observable properties of the CGM using smoothed particle hydrodynamics
(SPH)-based cosmological simulations.  In \citet{for13}, we examined
the general physical properties of the CGM, as well as absorption
statistics as a function of impact parameter, for \HI, \ion{Mg}{ii},
\ion{Si}{iv}, \ion{C}{iv}, \ion{O}{vi}, and \ion{Ne}{viii}.
In our simulations, the physical conditions traced by any
given ion are often a monotonic function of its ionisation potential,
such that low ionisation potential ions (which we will call low ions)
generally trace high-density gas very close to
galaxies at photo-ionisation temperatures (around $10^4$K), 
higher ionisation potential ions (which we will call high ions)
trace increasingly more diffuse, lower-density gas, and
the highest ions can trace hot gas when present (typically in halos
with masses $\ga 10^{12} M_\odot$).  We presented predictions for the
integrated column density and column density distributions for these
absorbers, as a function of impact parameter, and showed that 
low ion absorbers increase in strength dramatically when lines of sight pass
close to galaxies, while higher ions show a more modest increase closer to
galaxies.  These trends can be tested quantitatively
against current and forthcoming observations; comparisons that we are conducting now and will present in future work.

In this paper, we continue our focus on the physical state of the
CGM by examining the dynamical state of the CGM gas and its impact
on observable absorption line properties.  In particular, we ask
the question: Can we distinguish inflowing, outflowing, and ambient
gas based on CGM absorption signatures?  Here we focus on examining
this at low redshifts ($z\sim 0-0.25$), in anticipation of exploring
the constraints on models enabled by COS-Halos and similar CGM
projects; we leave an examination of this question at high redshifts
for future work.

To explore the dynamical state of CGM baryons, we implement
an analysis scheme that tracks gas as it moves in and out of galaxies.
Such tracking, we note, is uniquely enabled by our particle-based
simulation methodology.  By tracking exactly where the gas originates
and where it will eventually reside, we can
definitively identify which gas is inflowing, outflowing, and ambient
over a certain timescale.  Moreover, this enables us to track
material that once flowed out of a galaxy but that will eventually return to
a galaxy,  a process we call
{\it wind recycling}~\citep{opp10}. This component has key observational
implications becase it represents inflowing material that is enriched
and hence can be traced in metal absorption.  Indeed, we will
show~\citep[as in][]{opp10} that wind recycling provides a dominant
inflow contribution at the present cosmic epoch, which is by default
lacking in models that do not include outflows.
We will further show that absorption from low ions typically
comes from recycling inflows, while absorption from high ions
comes mostly from ancient outflows.

We organise our paper as follows: in \S 2 we introduce our simulations
and methods, in \S 3 we define and explore various categories of
inflowing, outflowing, and ambient material, and in \S 4 we present
observational diagnostics to distinguish among these categories.
In \S 5 we examine the physical conditions of these categories,
in \S 6 we discuss numerical considerations in our (and alternate) simulation methods, and in \S 7 we present our conclusions. 

\section{Simulations \& Analysis}
\subsection{The Code and Input Physics}
We use our modified version \citep{opp08} of the N-body+entropy-conserving smooth particle
hydrodynamic (EC-SPH) code \gad~\citep{spr05}, which is more fully described
in \S2.1 of \citet{dav10}. Our main simulation for this work is identical to
that in \cite{dav13}, and we
refer the reader to that work for a more detailed description.

We assume a $\Lambda$CDM cosmology \citep{hin09}: $\Omega_M = 0.28$,
$\Omega_\Lambda = 0.72$, $h={\rm H}_{o}/(100 \kms Mpc^{-1})=0.7$,
a primordial power spectra index $n$ =0.96, an amplitude of the mass
fluctuations scaled to $\sigma_8$=0.82, and   $\Omega_b = 0.046$.   We use a
cubic volume of 32${h}^{-1}$ Mpc on a side with ${512}^{3}$ dark matter and
${512}^{3}$ gas particles, and a softening length of $\eta=1.25h^{-1}$ kpc
(comoving, Plummer equivalent). The gas particle mass is 
$4.5\times{10}^{6}M_\odot$; dark matter particle mass is $2.3\times{10}^{7}M_\odot$.
 The stellar component of a Milky Way mass galaxy is thus represented with $\approx$ ${2}\times{10}^{4}$ particles. 

We incorporate cooling processes using primordial abundances as described by 
\cite{kat96}, with metal line cooling  based on tables from \cite{wie09} that 
assume ionisation equilibrium in the presence of the \cite{haa01} background.  
Star formation follows a \cite{sch59} Law calibrated to the \cite{ken98} 
relation, following \cite{spr03}. The ISM is modelled using the sub-grid recipe 
of \cite{spr03}, where a gas particle above a density threshold of $n_H=0.13$~cm$^{-3}$
 is modelled as a fraction of cold clouds embedded in a warm ionised 
medium following \cite{mck77}. We use the \cite{cha03} initial mass function 
(IMF) throughout. We account for metal enrichment from Type II supernovae, Type 
Ia SNe, and AGB stars, and we track four elements (C,O,Si,Fe) individually, as 
described in more detail in \cite{opp08}.

This simulation includes galactic outflows, which are implemented using a Monte 
Carlo approach. These outflows are tied to the SFR, $\dot{M}_{wind} = 
\eta\times SFR$, where $\eta$ is the mass loading factor. For this work we 
focus on the hybrid energy/momentum driven winds, or ``ezw" model. In the ezw 
model, the wind speed and mass loading factor depend on the galaxy velocity 
dispersion $\sigma$:

\begin{eqnarray}
\vw &=& 3\sigma\sqrt{{\rm f}_{L}-1}\\
\eta&=&\sigma_{o}/\sigma, {\rm\ if\ } \sigma > 75\kms\\
\eta&=&(\sigma_{o}/\sigma)^{2}, {\rm \ if\ } \sigma < 75\kms
\end{eqnarray}

where ${\rm f}_{L}$ is the luminosity in units of the Eddington
luminosity required to expel gas from a galaxy potential, $\sigma_{o}
= 150 \kms$, and $\sigma$ is the galaxy's internal velocity dispersion
\citep{opp08}. 

These scalings roughly capture the behaviour in recent models of outflows from 
the interstellar medium by \cite{mur10} and \cite{hop12}. We note that the
scalings for this ``ezw" model, and those in previous work by our group 
with the momentum-driven or ``vzw" 
model \citep[e.g.][]{opp08,dav10,dav11a,dav11b,opp12,for13}, which follow the 
scalings of \cite{mur05} are identical for higher mass systems. It is only in 
lower mass systems, i.e. those with $\sigma$ $<$ 75 $\kms$, where
the ezw model differs 
from the vzw model. In this particular simulation, we also add an artificial 
quenching mechanism in higher mass halos, as described by \cite{dav13}.

\subsection{Generating spectra with {\sc Specexbin}}

We use {\sc Specexbin}, described in more detail by \cite{opp06}, to calculate 
the physical properties of the gas. {\sc Specexbin} averages physical 
properties of the gas along a given sight line, then uses look-up tables 
calculated with {\sc CLOUDY} \citep[{}][version 08.00]{fer98} to find the 
ionisation fraction for the relevant ionic species. We use the same version of 
{\sc Specexbin} as in \cite{for13}, which includes a prescription for self-shielding from the ionisation background. This prescription results in a density threshold of approximately 0.01 ${\rm cm}^{-3}$ above which \ion{H}{i} is fully neutral and moves all the magnesium into \ion{Mg}{ii}. 
See Figure 1 of \citet{for13} for an example of a simulated spectrum. 

We fit Voigt profiles to the absorption features using \autovp\ \citep{dav97}.  
As in \citet{for13},  we consider all 
components within $\pm$ 300 $\kms$ to be associated with a galaxy, and we 
combine any components into systems with $\Delta v < 100 \kms$. We apply the 
same column density limits as in \citet{for13}, ${10}^{16}~{\rm cm}^{-2}$ for 
\ion{H}{i} and ${10}^{15}~{\rm cm}^{-2}$ for metal lines. All lines stronger than
this are set to this value to avoid having a single large saturated absorber 
skew the results, since such high column absorbers generally have highly 
uncertain column densities from Voigt profile fitting. As we showed in \cite{for13}, very weak lines are also poorly constrained given our assumed S/N=30, so we do not consider lines weaker than 30 m\AA.

For this work, we focus on targeted lines of sight (LOS). We randomly
select central galaxies in two different halo mass bins: ${\rm
10}^{10.75-11.25} \msolar$ (labelled ${\rm 10}^{11} \msolar$) and
${\rm 10}^{11.75-12.25} \msolar$ (labelled ${\rm 10}^{12} \msolar$).
For \mlo~, we select 250 galaxies, while for \mhi~there are only
221 central galaxies in the simulation so we use all of them in our
sample.  We choose impact parameters ranging from 10~kpc out to
300~kpc, with the spacing increasing slightly with impact parameter.
As in \cite{for13} we produce four LOS per galaxy at a position
{\it x}, {\it y}:  x+$b$, x-$b$, y+$b$, y-$b$, for a total of 1,000
LOS per $b$ per mass bin (844 for \mhi). We do not present results
within 10~kpc of galaxies since we cannot resolve the detailed
internal structure of the interstellar medium (ISM), and in any
case we are more interested in probing CGM gas in absorption towards
background objects that rarely lie at such small impact parameters.
For parts of this work we restrict our study to gas within the
virial radius of a central galaxy.  We define the virial radius
as the radius enclosing the virialization overdensity for our assumed
cosmology, as described 
in~\citep[see their eqs. 1 and 2]{dav10}, which at $z=0.25$
corresponds to roughly 90 times the critical density.
The median central galaxy stellar mass is $5.89\times10^8$ and $3.63\times10^{10}$
$M_\odot$ for our \mlo~and \mhi~ halos, respectively.

We look at \ion{H}{i}, \ion{Mg}{ii}, \ion{Si}{iv}, \ion{C}{iv},
\ion{O}{vi}, and \ion{Ne}{viii}
\citet{for13}. We select these ions as they are some of the most 
commonly observed species in the low-redshift CGM, spanning a wide range
of ionisation potentials. All the metal lines have 
doublets, making their identification in observed spectra more straightforward. 

\section{Inflows and Outflows}

\subsection{Identifying Inflowing and Outflowing Gas}

To examine the dynamical state of CGM gas, we must first
determine whether a given gas particle is inflowing, outflowing,
or ambient.  This is not trivial, because the constant cycling
of baryons within halos can conflate inflowing, outflowing, and
ambient gas.

Our approach is to make these divisions based on information available
in the simulations: the particles' past history and future fate. For this, we 
require two additional pieces of information from the
simulations:  The future location of particles, and the time when each wind 
particle was ejected.  

The first piece of information is the particles' future location.
For this, we cross-correlate all the non-ISM particles in our
simulation at $z=0.25$ with their location at $z=0$,
which is 3~Gyr later.  If during this time a
particle has been accreted into the ISM of a galaxy, or has turned into a star, or has been cycled through the ISM and subsequently ejected in an outflow, we identify it as
an ``accreting" particle.  {\it Accretion is thus defined as any gas that
was not in the ISM at $z=0.25$, but is either in or has passed through
the ISM or has turned into a star by $z=0$. } By definition, the ISM consists of all gas with $n_H\ge0.13$~cm$^{-3}$, the density threshold at which star formation is allowed. 

Next, we need the time of ejection for each wind particle.
We record each wind ejection event during the simulation so that for any
given time we can determine the {\it wind
age} of a given particle. If a particle {\it is not 
accreting} but has a non-zero wind age, it is considered an ``outflow" 
particle.  Note that the wind age can be quite large
if a particle was ejected in the early universe, and hence being
an outflow particle under this definition does not necessarily imply
that it is currently on its way out from a galaxy.  For particles
ejected multiple times, we only record the most recent ejection.
Finally, we define ``ambient" gas as particles that will not accrete 
by $z=0$ and was never ejected in a wind before $z=0.25$.
Thus we have the following divisions:

\begin{enumerate}
  \item {\bf Pristine Accretion}.  This is any accreting gas
  that has never been ejected in a wind.  In general, this gas
  tends to have quite low metallicity since it has never been in
  the ISM of a galaxy. It can have non-zero metallicity, however, due to enrichment from tidal stripping or in-situ star formation at an earlier epoch. 

  \item {\bf Recycled Accretion}. This is all accreting gas
  that was once ejected in a wind at least once before $z=0.25$.

  \item {\bf Young Outflows}.  These are gas particles that are
  not in the ISM at $z=0.25$, are not accreting, and have been ejected from the 
galaxy 
  in a wind between 0 and 1~Gyr before $z=0.25$. We choose 1 Gyr as this is roughly the 
time a particle would take to leave the halo if it simply got kicked into the a 
wind, never scattered or slowed down due to forces other than gravity. For our cosmology, 1~Gyr before $z=0.25$ is approximately $z=0.36$.

  \item {\bf Ancient Outflows}. Same as young outflows, only
  particles ejected more than 1~Gyr ago (before $z=0.36$).

  \item {\bf Ambient}. Gas particles that are not in the ISM
  of a galaxy at $z=0.25$, are not going to accrete onto a galaxy by $z=0$ and that have never been in a wind by $z=0$.
\end{enumerate}

All gas particles that are not in the ISM at z=0.25 are in one of these five categories. Note, in particular, that the distinction between outflow (AO and YO) material and recycled accretion material, and the distinction between ambient (AMB) material and pristine accretion (PA), is whether the gas will accrete by $z=0$. 

Note that young and ancient outflows are 
subdivided by time since ejection, but in general ancient outflows end up 
farther away from galaxies, while outflows ejected more recently
tend to stay closer.  This reflects the ``outside-in" IGM enrichment 
scenario occurring in our simulations as described in \citet{opp12}.  Also, 
recycled accretion can
in principle escape the halo before re-accreting.

\clearpage
\begin{figure*}
\centerline{ \includegraphics[width=6.3in]{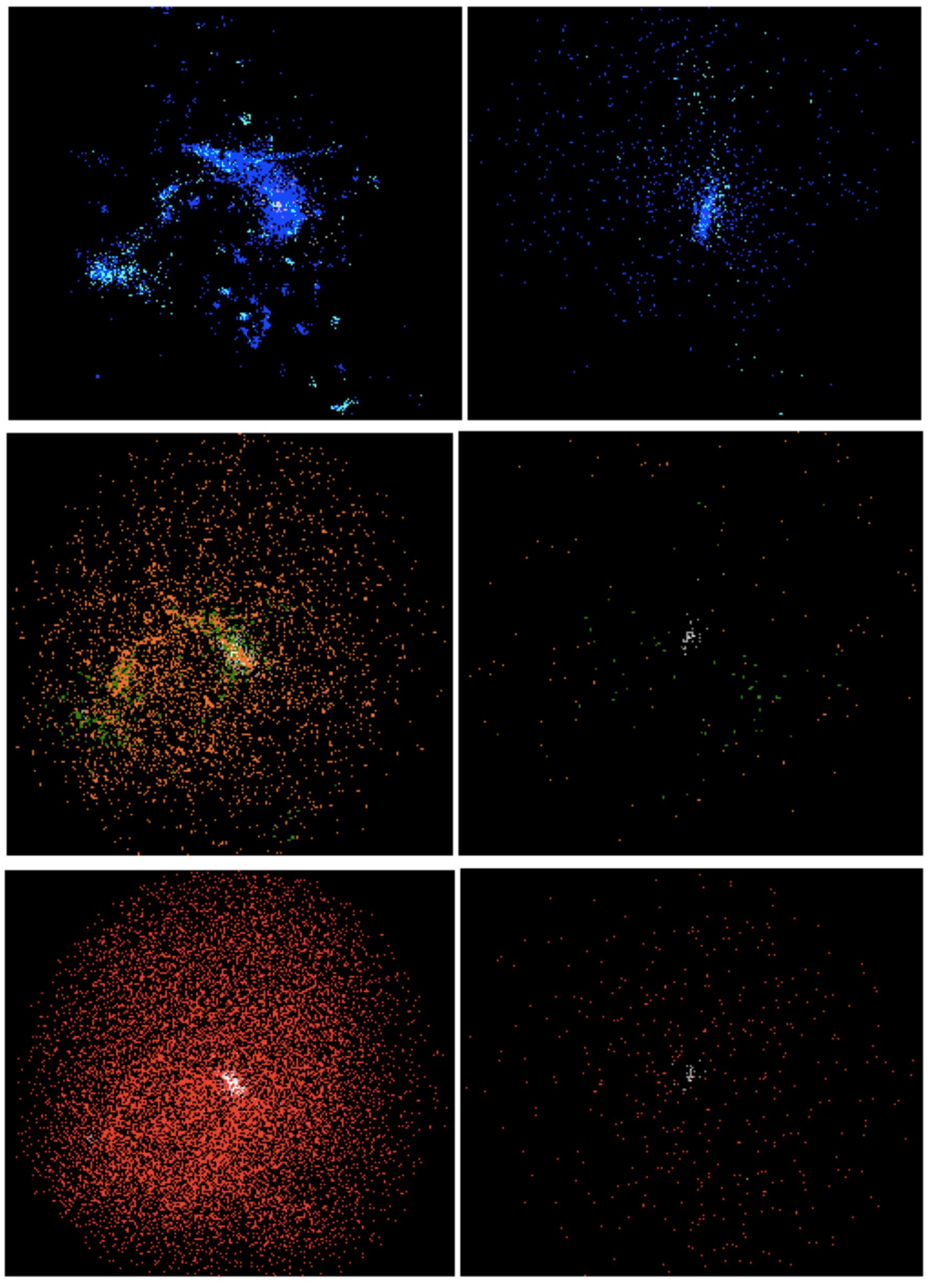}}
	\centering
 \caption{Left panels: SPH particles around a galaxy with halo mass $M_{halo}~=$\mhi~and a panel
width of 525~physical kpc. Right panels: SPH particles around a galaxy with halo mass
$M_{halo}~=$\mlo, panel width is 225~kpc.
Top panels show recycled accretion (dark blue), and pristine accretion
(light blue). Middle panels show young outflows (green) and ancient outflows
(orange). Lower panels show ambient gas (red). All panels also include stars (white).}
 \label{pretty}
 \end{figure*}
 \clearpage

We visualise these categories in Figure \ref{pretty}, which shows
a typical galaxy in a \mhi~halo (left panels) and in a \mlo~halo
(right panels).  We plot all the gas within the virial radius, 262 physical kpc
for the \mhi~halo and 112 physical kpc for the \mlo~halo. Both
galaxies are shown approximately edge-on. In each panel we plot the stars
in white. In the top panels we plot the accreting material
with recycled accretion in dark blue and pristine accretion in
light blue. In the middle panels we plot recent outflows in green
and ancient outflows in orange. In the lower panels we plot the ambient
gas in red.

These images reveal some qualitative trends regarding these categories.
Let us first examine the geometry. One may
expect accreting material to flow in along the filaments, parallel
to the disk of the galaxy, and outflowing material to come out
perpendicular to the plane of the disk.  However, the real picture is
not so simple.  First, only pristine accretion is expected to be
along the filaments while recycled accretion, having been ejected
from a galaxy, tends to be less confined to filaments. Second,
the filaments and disks are not always aligned
\citep{dan12}, so even pristine material flowing
in along a filament may not be parallel to the galaxy.  For both
masses in Figure \ref{pretty}, neither recycled nor pristine accretion
is clearly coming in along a well-defined axis. For the \mhi~halo,
the  recycled accretion appears to follow some sort of structure
while for the \mlo~halo it is more diffuse. 

Moving to the middle panels we see that in the \mhi~case the young
outflows generally are close to the main galaxy, its satellite, or the bridge that connects them. In the \mlo~ case they are more diffuse. Interestingly,
in neither the \mhi~or \mlo~case do the young outflows show an
obvious polar axis preference, as one might have expected, despite the fact that
the material is ejected in the direction of {\bf
v}$\times${\bf a}.  In part this is because our ``young" outflows
are not that young, and in many cases may already be rejoining the
accretion flow as recycled winds.  For both masses ancient outflows
are diffusely
distributed.  The ambient gas in the bottom panel fills
the halo roughly spherically, without a preferred direction.  We
will show that much of the ambient material is in a hot hydrostatic
gaseous halo, which is much more prominent at \mhi~than at \mlo~\citep{ker05,gab12}.

\begin{figure}
 \includegraphics[width=3.5in]{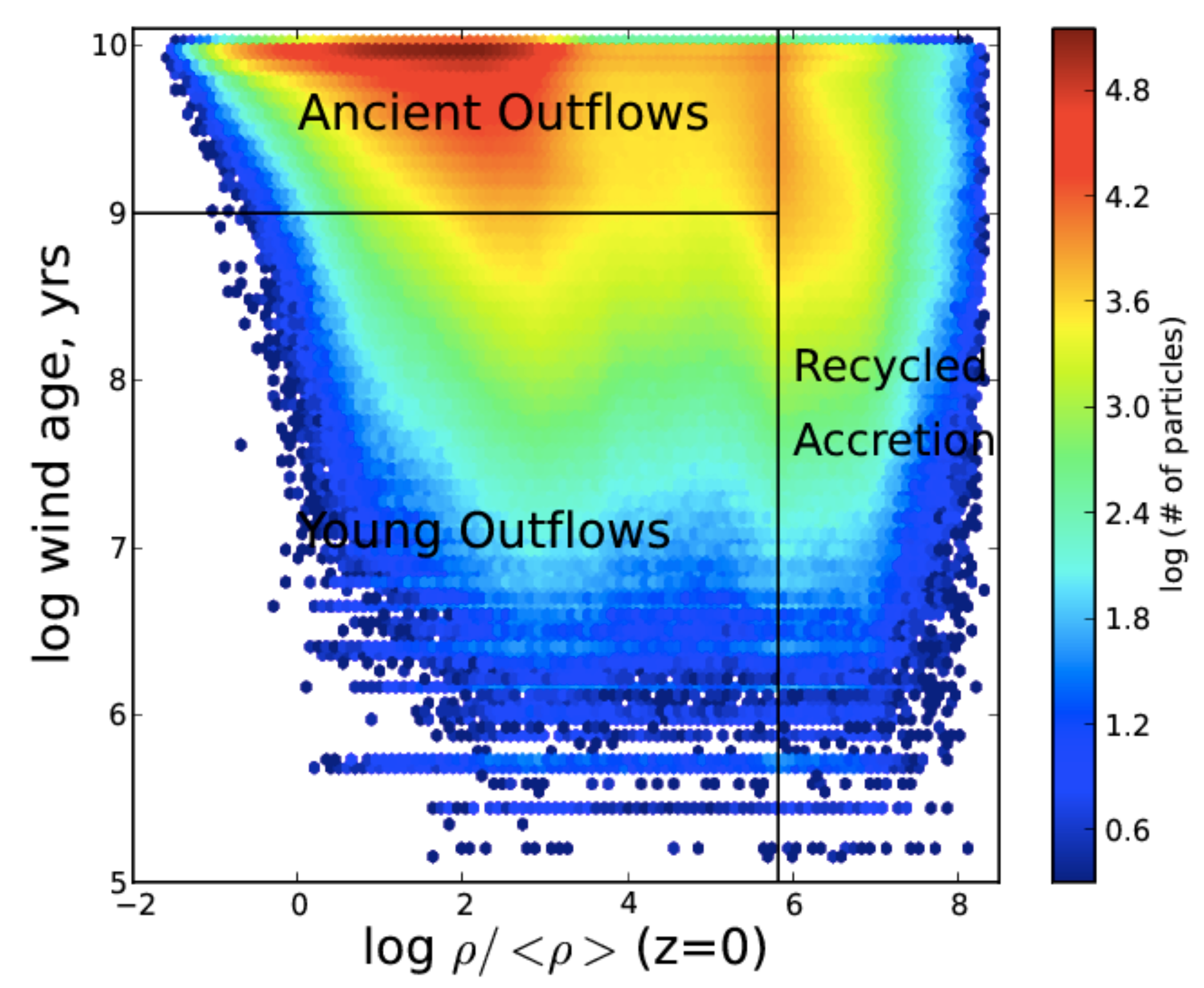}
\caption{The time (relative to $z=0.25$) since last wind ejection vs. density at 
$z=0$ of all gas not in the ISM at $z=0.25$. The vertical line denotes ISM 
densities at $z=0$. Gas to the right of the line is defined as accreting, because it will join the ISM of a galaxy by $z=0$. Gas to the left is defined as not accreting, unless it forms a star or gets launched into a wind (not shown on this plot).
The horizontal line at 9.0 shows the separation of young from ancient 
outflows.}
\label{div}
\end{figure}

To visualise these categories more quantitatively, in Figure \ref{div}
we show the wind age versus the $z=0$ overdensity of all non-ISM gas
particles (at $z=0.25$) that have ever been ejected in an outflow.
The vertical line demarcates the ISM overdensity threshold at $z=0$, i.e.
$n_H=0.13$~cm$^{-3}$.  Particles with densities greater than the
ISM threshold at $z=0$ are considered to be recycled accretion.
Particles that are not in the ISM at $z=0$ are subdivided by their wind age into
ancient outflows (wind age $>1$~Gyr) and young outflows 
(wind age $\le 1 $~Gyr).
The other two categories, pristine accretion and ambient,
do not appear in this plot since they have never been in a wind by definition. This figure shows not just how we have divided our categories but also the distribution of the gas particles. One can see there are more particles in the ancient outflow category than in young outflows, and that the number of particles in the recycled accretion category is large. We will quantify this in greater detail in later sections.

We emphasise that we do not {\it just} look at the $z=0$ ISM densities
(as shown in Figure \ref{div}) to determine whether or not a particle will
accrete by $z=0$. As we explained earlier, if a particle turns into a star or
enters the galaxy and is subsequently ejected we also consider that particle
as being accreted. In Table \ref{table:fatein} we show, in three halo mass bins, what fraction of
the accreting mass at z=0.25 will by z=0 end up in stars, in the 
ISM, or be ejected in a wind. In some cases, between $z=0.25$
and $z=0$, a particle is ejected multiple times and so at $z=0$ is again
at ISM densities; in those cases we give a preference to the ISM. For \mlo~halos, most of
the accreting mass ends up in the ISM, while for $10^{11.5} M_\odot$~halos just under half ends up in ISM. For \mhi~halos, the
distribution is almost completely evenly split amongst stars, ISM,
and wind ejection. The contribution of star formation and
wind ejection 
to the mass budget of accreting material is significant. Separately tracking
wind ejection and star formation events also gives us a finer time resolution
in the gap between $z=0.25$ and $z=0$. Note that Table \ref{table:fatein} does not imply that galaxies have 50\% gas fractions, as most stars formed from gas accreted at $z>0.25$.

\begin{table}
\caption{Fate of the accreting material}
\begin{tabular}{lccc}
\hline
log(Halo Mass) & 
Star &
Wind Ejection &
ISM
\\
\hline
\multicolumn{4}{c}{}\\
11.0 & 10\% & 18\% & 72\% \\
11.5 & 23\% & 30\% & 47\% \\
12.0 & 31\% & 35\% & 34\% \\
\hline
\end{tabular}
\label{table:fatein}
\end{table}

\subsection{Mass Budgets}

\begin{figure*}
 \includegraphics[width=7in]{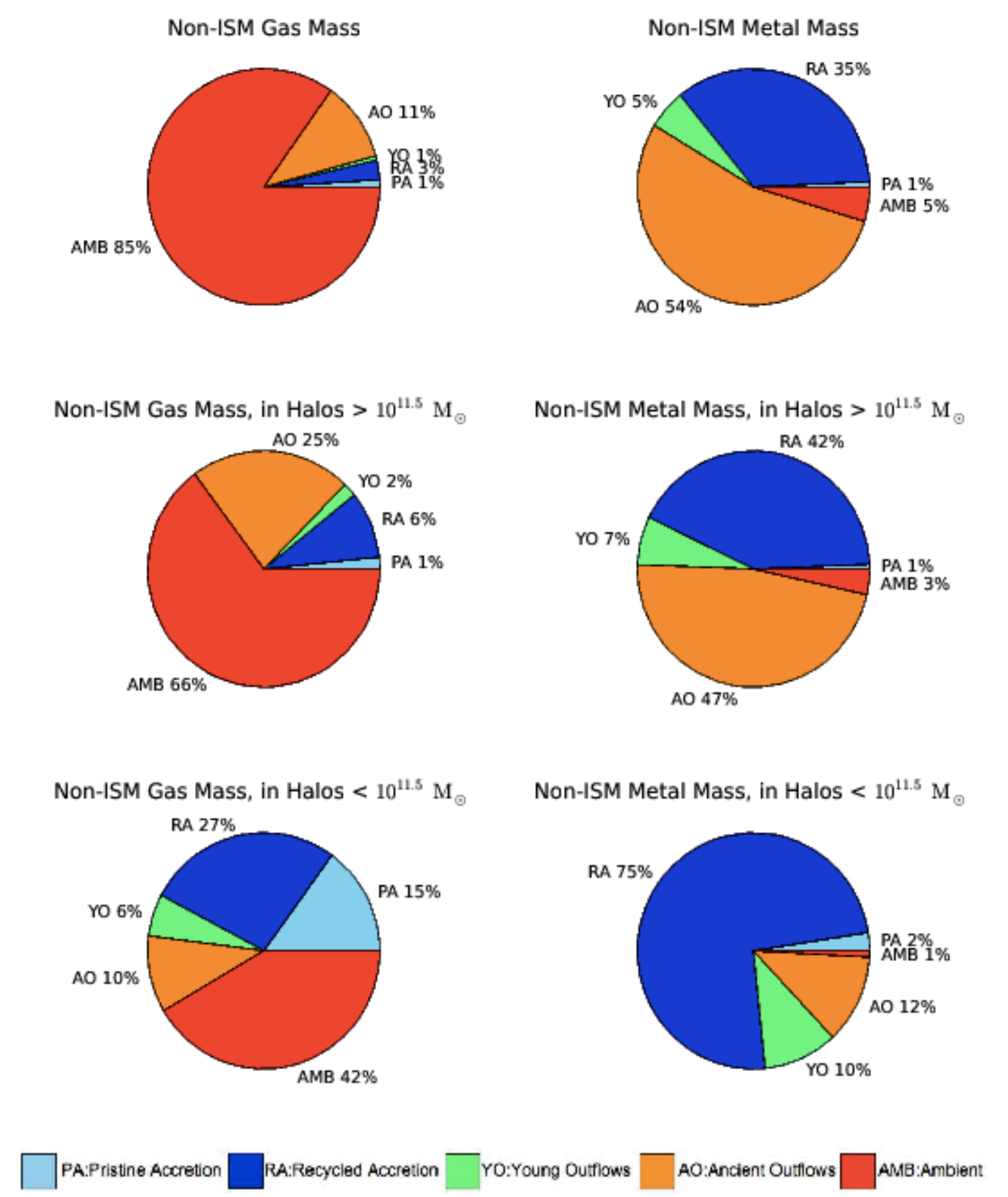}
\caption{The non-ISM gas mass budget (left) and gas-phase metal
budget (right) at $z=0.25$.  Upper plots show the full simulation,
and lower plots show only gas within the virial radius of a halo
at $z=0.25$, for high (middle) and low (bottom) mass halos. Ambient is gas that has never been in a wind, before
or after $z = 0.25$, that is not accreting.  Ancient outflow is for non-accreting gas particles ejected in a wind more than 1~Gyr before $z=0.25$. Young outflow is for non-accreting gas particles ejected in a wind $\le$ 1~Gyr before $z=0.25$. Recycled accretion is gas that has been ejected in a wind by $z=0.25$, and will join or pass through the ISM of a galaxy or become a star by $z=0$.  Pristine accretion is gas that has not been ejected in a wind by $z=0.25$ and will join or pass through the ISM of a galaxy or become a star by $z=0$. }
\label{pie}
\end{figure*}

To begin analysing the gas in these various categories,
we first examine their mass fractions.  We do
this both for all the gas in the simulation and for gas within the virial
radius of dark matter halos, which we identify as CGM gas.
As we discussed in \S1,
there are various definitions for CGM in the literature, and in \citet{for13}
we argued for 300~kpc as an appropriate radius for the
CGM based on the metal absorption extent, at least for $\sim L^*$
galaxies.  However, we also showed that the extent of absorption
depends on the particular metal ion and there was not a clean
distinction between enriched and unenriched regions.  Hence, here
we will use a CGM definition that is perhaps more removed from
direct observations but is well-motivated and well-defined
theoretically, namely the virial radius.  Our definition for the virial radius is discussed in section \S~2. This is at least something
that is directly quantifiable and whose extent scales with the
galaxy mass.

Figure \ref{pie} shows pie charts for the total mass (upper left)
and metal mass (upper right) of all non-ISM gas particles in the
simulation at $z=0.25$, broken down by category, while the middle and bottom panels show
analogous pie charts considering only non-ISM gas within high- and low-mass halos, respectively. We choose $M_{halo}={10}^{11.5}\msolar$ as the dividing line between high- and low-mass halos because that is where we see a crossover between hot and cold gas fractions, which we discuss later. We define particles as within the halo if ${R}\le{R}_{\rm vir}$. 
(For reference, 
we note that ISM gas makes up 1.8\% of the total gas mass of the simulation at $z=0.25$.) 

In the upper left panel, we see that most of the non-ISM gas mass in the 
simulation consists of ambient material.  Even though outflows
in our simulations are ubiquitous and have mass loading factors
typically of unity or above~\citep{opp08}, the majority of baryons in
the Universe have never been in a wind since the baryonic fraction
in galaxies is small, only 7\% in this model.  Ambient gas,
much of which consists of diffuse gas in the IGM, accounts for
85\% of all baryons;
this material does not participate in the baryon cycle by $z=0$
since it has neither been in an outflow nor has it been accreted.
Ancient outflow is the next largest category at around 11\%, which is 
still larger than the fraction of baryons in stars (6\%), showing that the mass
in outflows exceeds that in stars globally, as noted in our earlier
simulations \citep{opp08}.  Young ($\le1 $~Gyr old) outflows
comprise only about 1\% of the mass, which is not surprising since the outflow 
rate roughly tracks the star formation rate, and this is much smaller at the
present epoch than at high redshifts.  The global accreting gas mass
fraction from $z=0.25-0$ is 4\%, of which the vast majority was
previously in a wind (3\%). Pristine accretion over the 3~Gyr
from $z=0.25$ to $z=0$ only accounts for 1\% of all baryons. Of the material that accretes between $z=0.25$ and $z=0$, 25\% of it (by mass) is pristine accretion; the other 75\% is recycled accretion.  

The story is quite different if one considers the metal mass (upper right pie
chart).  Now ancient outflows contain 54\% of all cosmic metals
and recycled accretion contains over one-third.  Young
outflows, which contain only 1\% of the total mass, still contain 5\% of
the metals.  Ambient material, having essentially never resided
inside a galaxy, contains a very small amount of metals relative
to its mass fraction, and pristine accretion likewise has a negligible
metal mass content.  The metals that are present in the ambient or pristine
accretion gas has three possible sources within our simulations: Type Ia 
supernova, AGB stars, or tidal stripping. The metals in pristine accreting
material mostly owes to tidally stripped ISM material, which then reaccretes. Pristine accretion has an average metallicity of $0.07 Z_{\odot}$, while recycled accretion has on average approximately solar metallicity. 
It is worth emphasising that overall accreting material is typically
significantly metal enriched, and hence does not usually have low metallicity
as is sometimes assumed, at least at these low redshifts. 

In the middle and lower panels of Figure \ref{pie} we restrict ourselves to
only non-ISM material within the virial radius of dark matter halos, i.e. the
CGM, which includes roughly 20\% of the total gas mass and 70\% of the total
metal mass of non-ISM gas particles.  
From the mass fractions
in the lower left panels, we see that the fraction of CGM material participating
in the baryon cycle is significantly larger for both low- and high-mass halos --  the inflow and outflow categories are more
prominent.  Nonetheless, ancient outflows still dominates over young
outflows, and recycled accretion dominates over pristine accretion. High-mass halos have a larger percentage of ambient material, since they can keep their gas hot via a stable virial shock~\citep{bir03} and prevent it from falling back in. The relative contribution of accretion in high-mass halos is smaller than in low-mass ones, also because of temperature: hot halos can prevent infall. 

The importance of ``halo fountains", i.e. recycled
accretion that never leaves the halo, is shown in the metal fraction
plot within halos (lower right).  For high-mass halos, almost half the metal mass within
the halos at $z=0.25$ will, by $z=0$, be accreted onto galaxies.
The remaining half of the metals are mostly in ancient outflows that are still
trapped or recaptured within halos via ``outside-in"
enrichment~\citep{opp12}, with young outflows having a slightly
increased metal proportion relative to their gas mass because outflows today
are somewhat more metal-rich than outflows at earlier epochs. This owes
to the upwards evolution of galaxies along the galaxy mass-metallicity relation in
these models~\citep{dav11b}. For the low-mass halos, the contribution from recycled accretion is dominant: fully 75\%. In these low-mass halos, the metals either escape the halo completely or fall back in, whereas in high-mass halos there is a hot halo that can keep the metals bound to the halo but not falling back in. 

\begin{figure}
 \includegraphics[width=3.4in]{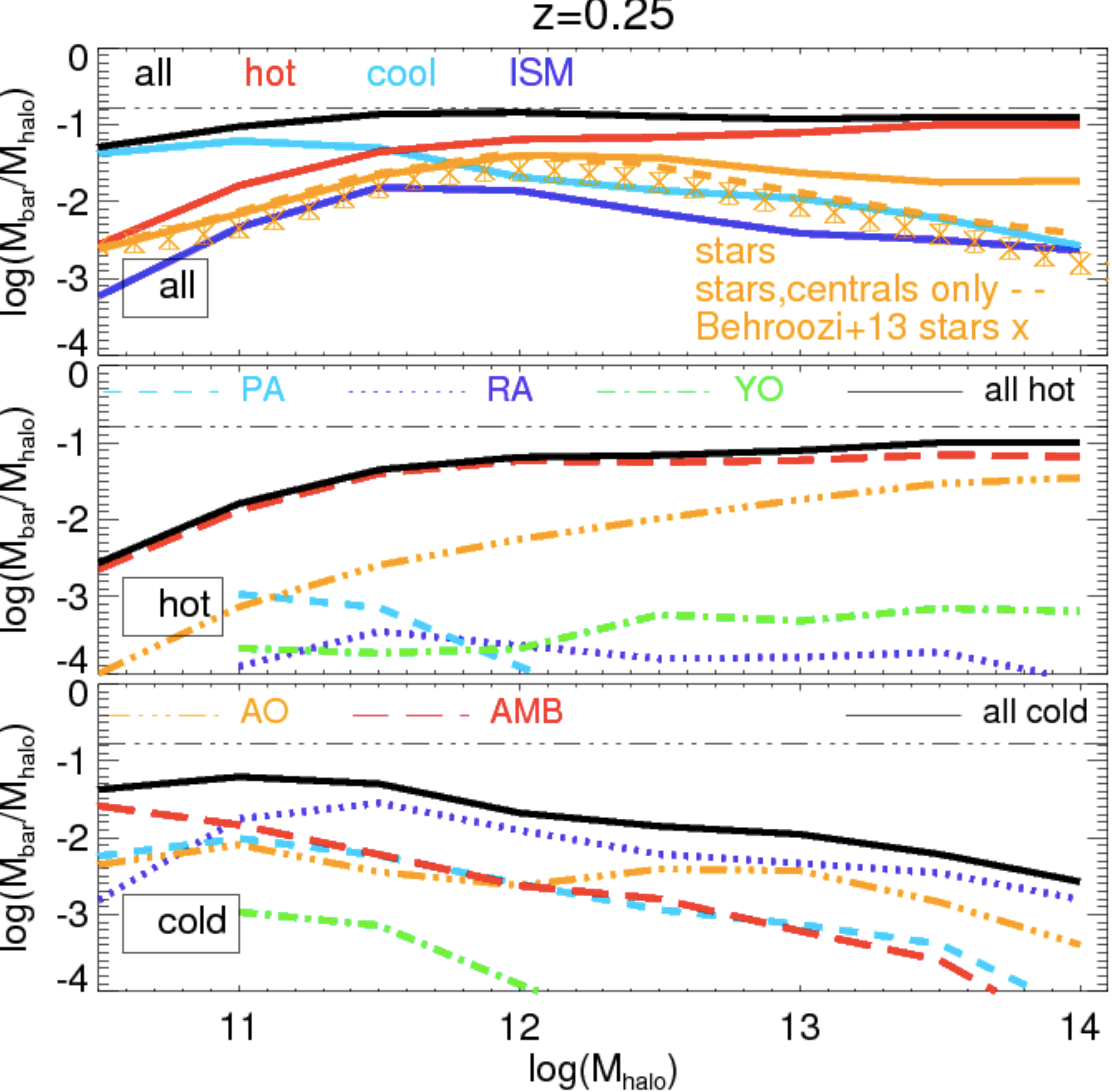}
 \caption{Upper panel: the fraction of baryons that are hot  
($>$ ${\rm10}^{5}$ K; red) ,
cool  ($<$ ${\rm10}^{5}$ K; light blue), the ISM (i.e. star-forming; dark blue), and in 
stars (orange) vs. halo mass at $z=0.25$. 
We plot the total baryon fraction in halos (solid black).  The black horizontal 
dot-dot-dot-dash line represents the global ratio $\Omega_b$/$\Omega_m$. 
We also plot the fraction of baryons in just the hot phase (middle panel) 
and the cold phase (bottom panel) vs. halo mass
divided into categories as labelled.}. 
\label{halofrac}
\end{figure}

In Figure~\ref{halofrac}, we investigate the CGM mass budgets further
by breaking down our inflow/outflow categories by their phase.  We
consider four phases: hot ($T>10^5$~K), cold ($T<10^5$~K), ISM ($n_H
\ge 0.13$~cm$^{-3}$) , and stars.  The top panel shows the fractional
CGM gas mass, relative to the total mass, by phase as a function
of halo mass.  The black line shows the total baryonic fraction,
which hovers around but is slightly below the universal baryon
fraction of 0.164, (dot-dot-dot-dash horizontal line), until it
starts to drop at $M_{\rm halo}\la 10^{11.5} M_\odot$.  The hot and
cool gas fractions cross over at roughly this mass, with the CGM
becoming hot gas-dominated at higher masses as is typically seen
in these types of simulations~\citep[e.g.][]{ker05,gab12,nel13}.
(This crossover also serves as justification for the mass breakouts
shown in Figure \ref{pie}.) The stellar and ISM gas fractions have
a peak at $M_{\rm halo}\approx 10^{12} M_\odot$; star formation is
suppressed at lower masses by increasingly strong outflows and at
higher masses by our quenching prescription.  The overall shape
agrees reasonably well with that inferred from halo abundance
matching studies at $z=0.1$~\citep[e.g.][]{beh13}. As shown in
\cite{beh13}, there is very little evolution in ${\rm M}_{\rm
bar}/{\rm M}_{\rm halo}$ as a function of ${\rm M}_{\rm halo}$ from
$z=1$ to $z=0.1$, so comparing this work's $z=0.25$ data to
\cite{beh13} at $z=0.1$ is acceptable. There are some differences
at higher masses, because we include satellites in the stellar
content of our halos while \citet{beh13} does not include them. As
shown in \cite{ber03}, satellites make a larger contribution to the
stellar mass budget at larger halo masses.  If we exclude satellites
(orange dashed line), we get much better agreement with \citet{beh13}.
This is essentially equivalent to saying that the stellar mass
function in this model agrees well with the observations, which
was shown directly by~\citet{dav13}.

In the middle and lower panels we consider the hot and cold (non-ISM)
CGM components, demarcated at $10^5K$, and split them in our inflow/outflow
categories as labelled.  The black lines show the total hot (red line in
upper panel) and total cool (light blue in upper panel) fractions for reference.
The hot phase is completely dominated by ambient gas at all masses.
This indicates that hot halo baryons are generally not participating
in the baryon cycle, at least in our simulations.  Ancient outflows are approximately an order
of magnitude smaller in mass fraction, except at the largest masses
where the outflows are more efficiently retained or re-accreted
within group potentials and rise to become half the ambient gas mass.
The other categories provide an essentially negligible contribution to the hot gas.
We note that in our models we eject the winds at the ISM temperature, which are
then allowed to interact with the CGM gas once it escapes the ISM.
However,  given that it is metal enriched it tends to cool quickly even
if it shock-heats initially.  This enhances the recycling 
of that material, and hence hot halo gas contains little of this
component.

By contrast, the cool phase has more mass undergoing baryon
cycling.  For halos with $M_{\rm halo}>10^{11}M_\odot$, recycled
accretion dominates the mass budget, owing to short recycling times
in massive halos~\citep{opp10}.  For smaller mass halos, ambient gas again
dominates, and pristine accretion becomes the most important accretion
channel.  This transition from enriched to pristine accretion in the
cool gas could provide an interesting signature of recycling in
absorption lines.

To recap, we divide all non-ISM gas at $z=0.25$ into five
categories: {\bf pristine accretion}, which holds a small fraction
of both the total gas mass and the metal mass but dominates accretion in dwarf
galaxy halos; {\bf recycled accretion}, which dominates accretion
for sizeable halos and holds a substantial fraction of their metal
mass; {\bf young outflows}, which are a sub-dominant component of
mass and metals in the CGM, {\bf ancient outflows}, which hold roughly
half the total metal mass in the CGM; and {\bf ambient material},
which contains a small fraction of the metals but a majority of the
gas mass, particularly the hot gas mass, both globally and within the CGM.  

\subsection{Relation to Dynamics}

We have defined inflows and outflows based on their future and past
history relative to $z=0.25$.  It is interesting to see how
these definitions relate to the actual dynamics of the gas at
$z=0.25$.  One expected trend would be that outflows, particularly young
outflows, should have an outward radial velocity relative to the
galaxy and that inflowing gas should have a negative radial velocity.
To examine this, we calculate the radial velocity ${\rm v}_{\rm r}$ of gas
particles
relative to the nearest central galaxy, and compare it to the halo
virial velocity ${\rm v}_{\rm vir}$.

\begin{figure}
\includegraphics[width=3.4in]{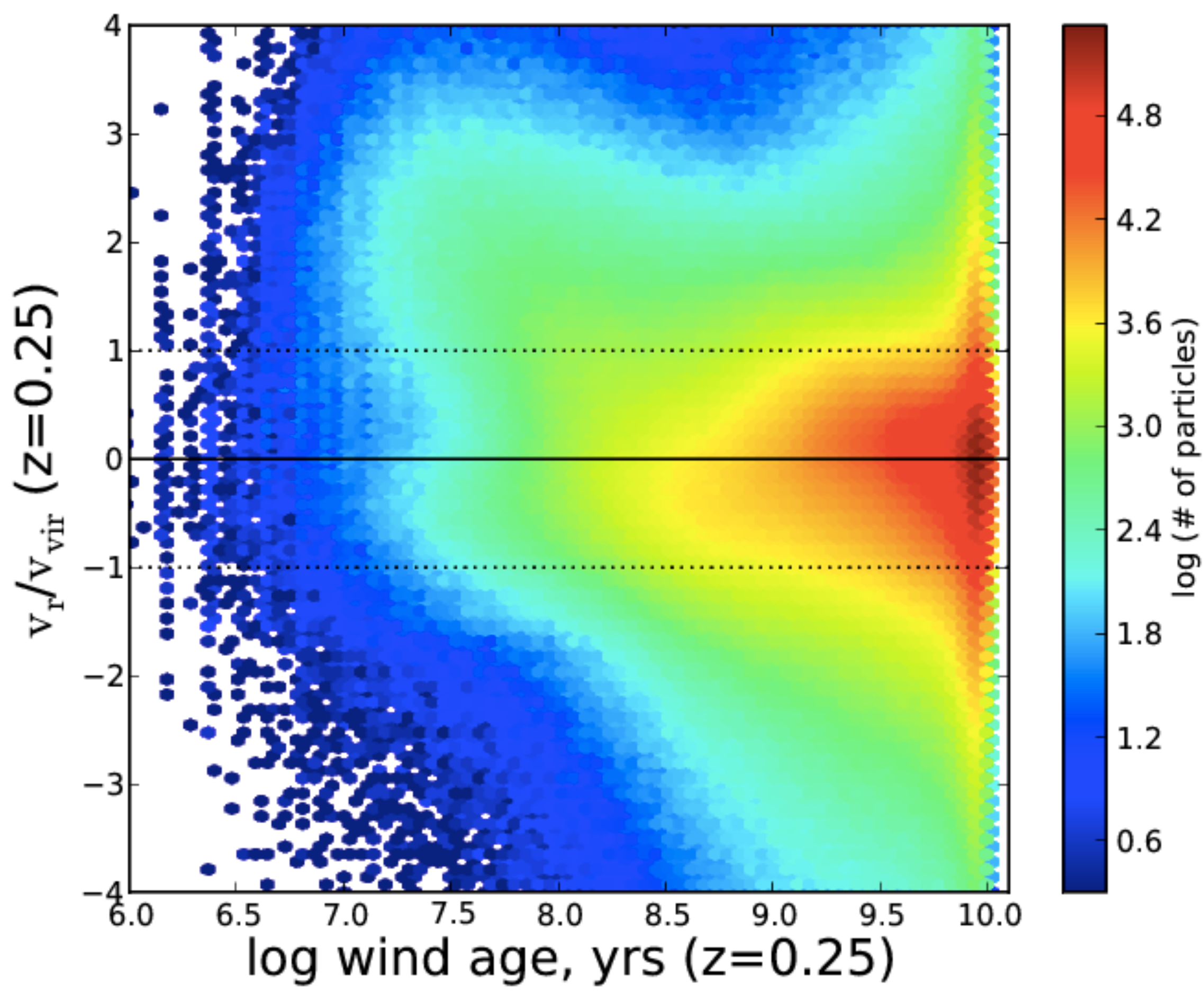}
\caption{${\rm v}/{\rm v}_{\rm vir}$ vs. wind age for all non-ISM gas particles in the simulation at $z=0.25$}
\label{vcomp}
\end{figure}

Figure \ref{vcomp} shows ${\rm v}_{\rm r}/{\rm v}_{\rm vir}$ vs. wind age
for all non-ISM particles in the simulation. For ages in the range of roughly ${10}^{7}$ to ${10}^{9}$ years, our
dividing line between young and ancient outflows, there is a clear
asymmetry in this distribution towards positive ${\rm v}_{\rm r}$, i.e. outflowing
gas, which is much more pronounced for more recently ejected winds.
A more subtle trend is that, excluding the gas that is in the
outflowing ``plume" to small age and large ${\rm v}_{\rm r}$, there is a
negative velocity for the remaining wind gas.  This corresponds
to wind recycling, which we will explore in more detail below.
Despite the presence of clearly outflowing gas, even recent outflows
can span a range of velocities, including negative.  This indicates
that instantaneous radial velocity alone is not a robust identifier of relatively
recent ejection.

Another approach to examining how our adopted categories relate to
the dynamics of the gas is to look at velocity-radius plots.  In such
plots, the outflows would be expected to have a large positive radial
velocity close to galaxies, diminishing as it moved further away,
while inflows might be expected to show up as a net negative velocity,
perhaps increasing as one approaches a galaxy.

\begin{figure*}
 \includegraphics[width=6.7in]{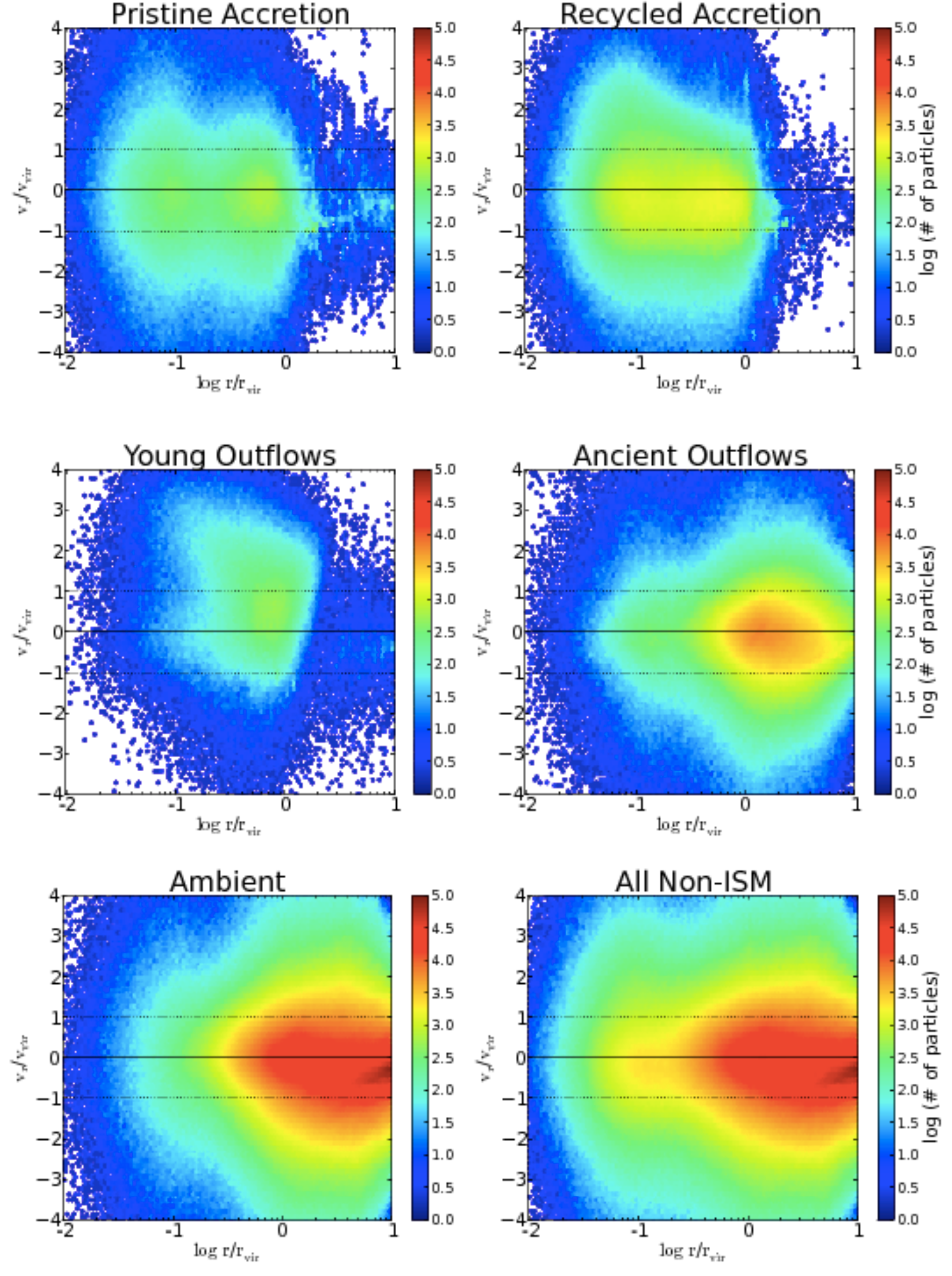}
\caption{V $\cdot$ r/abs(r) (divided by the virial velocity) vs. radialdistance from the host galaxy (divided by the virial radius), for non-ISM gas at $z=0.25$ divided into each of the categories as labelled. To guide the eye, we have added solid horizontal lines at ${\rm v}=0$, and dotted horizontal lines at ${\rm v}=\pm {\rm v}_{\rm vir}$.}
\label{velcat}
\end{figure*}

Figure \ref{velcat} shows the radial velocity (scaled
to the virial velocity) versus radius (scaled to the virial radius)
split into our categories as labelled. We associate each gas particle with
the galaxy to which it is most bound. Because satellite galaxies can both
eject winds and accrete gas, which could confuse our analysis, in these plots
we restrict ourselves to gas particles associated with central galaxies. The scale for each panel is logarithmic in particle number, and
spans the same range in each panel for ease of comparison.
We note that each category does {\it not} have the same number of
particles.
The top two panels show accretion,
pristine (left) and recycled (right).  The middle two panels show
young (left) and ancient (right) outflows.  The bottom panels show
ambient gas (left) and the sum of all the categories, i.e. all non-ISM
gas (right).  The solid line indicates zero velocity and the dotted
lines delineate the virial velocity.

We begin by looking at all gas particles in the simulation, not in
the ISM at $z=0.25$ and associated with central galaxies (lower right panel). 
With the exception of an inflowing plume at large ${\rm r}/{\rm r}_{\rm vir}$, which we
discuss in more detail below, in general the distribution is fairly
symmetric.  The distribution is also fairly smooth: there is not a neat
``outflow" plume, a tight ``inflow" plume, or a distinct ``rotating with disk"
component. Rather, one sees a more ambiguous picture, and hence robustly deciding
which positive velocity gas to call outflows versus rotation or
ambient motion can be quite difficult.  
This is why we choose to make our
distinctions not based on the velocity, but in the manner described in \S
2.1

Looking at the different categories, pristine accretion (upper left) has
the peak of its distribution at or slightly less than zero, depending on
${\rm r}/{\rm r}_{\rm vir}$, indicating an  inflow.
There is also a slight inflowing plume
at large radius, as seen in the ``all non-ISM" panel. It is remarkable that
so much pristinely accreting material has positive velocities, and it shows
that accretion is not so easily identified as merely inward-moving gas. There is a lot of random motion in the halo, so even material that will eventually accrete (onto either a central or satellite galaxy) can be moving away from the central galaxy, as shown here. 
Recycled accretion (upper right) shows similar trends: the peak of its
distribution (yellow region) is shifted slightly below zero, but much of
the material has a positive velocity, especially at
radii $\approx 0.1{\rm r}_{\rm vir}$. This is recent wind
material, which will turn around and accrete by $z=0$.

The middle two panels show the outflows, young (left) and ancient
(right).  The young outflows show a distinct asymmetry towards
outflowing gas.  However, at any radius,
young outflows are both moving towards and away from the galaxy.  In
other words, there is not an obvious demarcation in radial velocity
that will uniquely isolate outflows occurring within the last $\approx$ Gyr.
One could demarcate the plume at high ${\rm v}_{\rm r}$ and low $r$, but this
will contain only a small fraction of the young outflowing material.
 Meanwhile,
the ancient outflows show, unsurprisingly, even less asymmetry,
with the exception of the the same inward plume seen earlier.

The ambient gas (lower left) shows essentially perfect symmetry about
${\rm v}=0$, as
one would expect for gas that is neither in a wind nor going to be
accreted.  However, there is a strong concentration within the
large radius plume, which is
seen in the other panels as well. This owes to massive
halos only ($>$ ${10}^{11.5}$ $\msolar$); lower mass halos do not have
this feature.  This plume simply owes to the
gravitational growth of structure, as gas falls into halos
at large radii.  Such gas gives rise to an infall signature at large
radii as observed at $z\sim 2$~\citep{rud13}.
However, once it falls into the halo, ambient gas
does not continue its motion towards the galaxy, but instead settles
into a virialised halo that has essentially no net velocity relative
to the galaxy.

In summary, the various categories of inflow and outflows show
distinct trends in their kinematics.  Accretion has a net
negative inflow, although some of that material has positive velocities.
Young outflows
have a distinct asymmetry towards having more outwards moving gas,
but there is still a wide range of radial velocities
in this category, even at small radii.  In terms of wind age, the
youngest outflows show a clear outflowing tendency, but this signature
mostly disappears for wind ages above a Gyr.  Such ancient winds show
little net infall or outflow from within the halo, and in many ways are
much
more like ambient material.  Generally, it is difficult to separate
these categories cleanly using cuts in velocity space, even when one includes
radial distance information.

\section{Observables}

With a better understanding of the dynamical state of the CGM in
hand, we now turn our focus to the observable signatures of accretion,
outflows, and ambient material.  In principle, a sufficient suite of
observed absorption line tracers could directly constrain both the
physical state and dynamics of the gas relative to a nearby galaxy.
However, in practice, absorbers around halos are sparse and are not
commonly visible in multiple metal tracers.  Together with the 
uncertainty over the multi-phase nature of CGM, it is 
challenging to extract robust information directly from the data.
Hence simulations such as ours can provide some insights into 
mapping specific ions onto gas with specific physical and dynamical
properties.

In \citet{for13} we focused on how absorbers trace the physical 
conditions of the gas, namely the density and temperature.  Here, 
we extend this to now consider the dynamical state of the gas,
and determine whether there is some reliable mapping between certain
absorption line tracers and whether gas is inflowing, outflowing,
or ambient.

\subsection{Column Density vs. Impact Parameter}

\begin{figure*}
\includegraphics[width=7in]{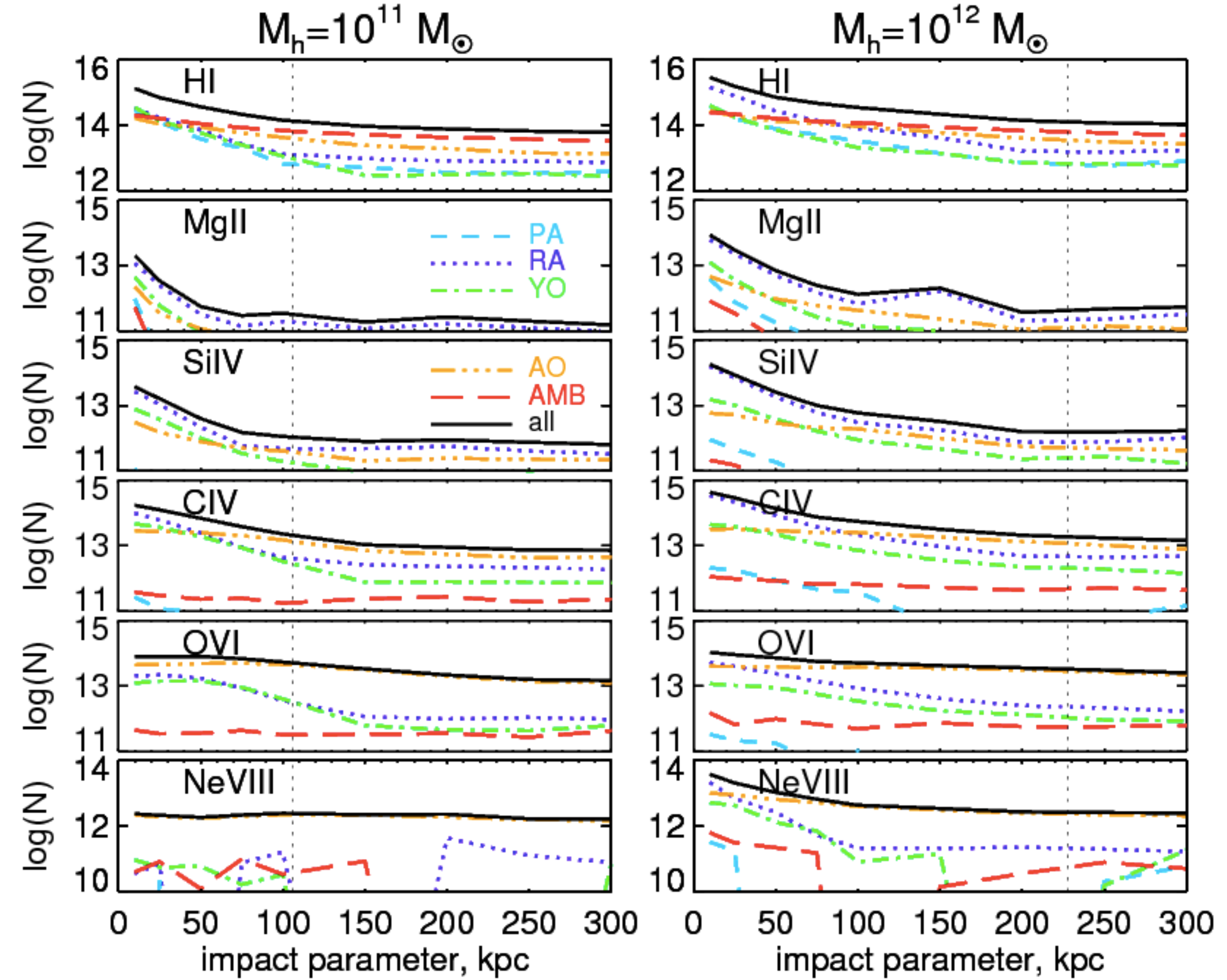}
\caption{The average column density per LOS for targeted lines of sight
around \mlo~(left panels) and  \mhi~(right panels) halos for different species
(as labelled). The black
solid line includes all the gas. Coloured lines indicate results from 
including only absorption from:
pristine accretion (light blue dashed),
recycled accretion (dark blue dotted), young outflows (green dot-dashed),
ancient outflows (orange dot-dot-dashed), and ambient (red long-dashed).
Vertical lines indicate the approximate virial radius at the 
midpoint of each mass bin. Note the yrange is not the same for each ion, though it always spans 4 dex.}
\label{NvsB}
\end{figure*}

In Figure \ref{NvsB} we plot the mean column density per line of
sight of different species (as labelled)
versus impact parameter for galaxies with a halo mass
of \mlo~(left panels) and \mhi~(right panels).
We start at 10~kpc, which is the typical extent of the ISM, and go out
to 300~kpc, roughly delineating the CGM as argued in \citet{for13}. 
We compute this by summing the column density within
$\pm 300$~$\kms$ of the galaxy (for all lines of sight), and then divide by the number of lines of sight.  
We choose to plot a mean versus a median to promote a fairer comparison with the data,
since a median is sensitive to the (often variable) detection threshold. 
We then subdivide the mean column density by inflow/outflow category.
In practice, we create a simulation snapshot containing only gas particles of
a given category, obtain absorption spectra using
{\tt Specexbin}~\citep{opp08}, and get
column densities by fitting Voigt profiles using {\tt AutoVP}~\citep{dav97}.  
The solid black line is for all the gas in the simulation, and the
coloured broken lines show the breakdown by category,
as indicated in the legend.  The solid black line is akin to what an observer would actually see, the coloured lines are what an observer would see if the universe consisted only of gas in the labelled category. 

Looking at HI absorption in $L^*$ halos (\mhi; top right panel), we see that
recycled accretion provides the
dominant contribution at impact parameters less than about 75~kpc
(all distances are physical). Beyond
that, ambient gas and ancient outflows make up the bulk of the
absorption.  In \mlo~halos (top left panel),
ambient gas dominates at all impact parameters greater than about 25~kpc, followed by ancient outflows. At very
small radii, recycled accretion also becomes important for low-mass halos.
Both theoretical and observational works have claimed that 
accretion can be probed
via high-column density \HI\ lines~\citep{fum11,vdv12,leh12} and our
results support this claim, although in our case we are probing lower column
density gas than the Lyman Limit systems (LLS) typically considered in those
works.  In our case, the typical column density at 25~kpc is
$10^{15}\cdunits$, which is well below a LLS but strong nonetheless.

For \ion{Mg}{ii}, the majority of absorption owes to recycled
accretion at all impact parameters and at both halo masses.
Outflowing material altogether provides a much lower contribution to the absorption; close in, young outflows are more prominent, while
beyond 50~kpc ancient outflows contribute more.  Low-metallicity
gas, i.e. pristine accretion and ambient gas, makes only a negligible
contribution.  At all impact parameters,
\ion{Mg}{ii} traces gas that will fall into galaxies by $z=0$.  Physically, gas in our simulations that is
sufficiently cold and dense to give rise to \ion{Mg}{ii} absorption
cannot support itself against gravity, and hence accretes on a dynamical
timescale. Thus our simulation indicates a tight connection between \ion{Mg}{ii} absorption and star formation that occurs within 1-2 dynamical timescales.

The story is similar for \ion{Si}{iv}: recycled accretion is the most important
source for absorption at all impact parameters and halo masses.  However,
it is less dominant than in the \ion{Mg}{ii} case.  Once again young
outflows dominate the outflow contribution close in, but the crossover
point with ancient outflows occurs farther out, at $\approx$ 50~kpc.  Hence
\ion{Si}{iv}, despite its rather high ionisation state, behaves
much like \ion{Mg}{ii} because the typical temperature
and density giving rise to \ion{Si}{iv} absorption is more similar
to \ion{Mg}{ii} than to that of higher ions~\citep{for13}.

For \ion{C}{iv} the behaviour changes somewhat and also becomes
more dependent on halo mass.  For \mhi\ halos at low impact parameters,
recycled accretion provides the dominant contribution to the total
absorption, but beyond about 100~kpc ancient outflows become the
largest contributor.  For \mlo\ halos, within 50~kpc, recycled accretion, young outflows,
and ancient outflows all are equally important,
while at larger impact parameters ancient outflows dominate, followed by recycled accretion
and then young outflows. Hence, depending on
the impact parameter and the halo mass, \ion{C}{iv} can trace recycled
winds, and either young or ancient outflows.

For the high ions \ion{O}{vi} and \ion{Ne}{viii}, recycled accretion
dominates at small impact parameters in large halos, otherwise ancient
outflows dominate.  Interestingly, the shape
of the black line for the high ions is flatter than for low or mid
level ions -- the strength of the absorption depends less on the
proximity to the galaxy, as noted by \citet{for13}.

Material that has never been in a wind (ambient or pristine accretion)
does not provide a significant amount of absorption in metal lines.
For \ion{H}{i}, however, ambient gas plays a major role, more than any other
category at radii $\ge$ 75~kpc. This is not surprising given the huge amount
of mass contained within the  ambient gas, as seen in Figure \ref{pie}. 
However, at radii $\le$ 75kpc, recycled accretion provides the dominant 
contribution for \ion{H}{i} and this region typically contains the
strongest \ion{H}{i} lines~\citep{for13}. Hence, recycled accretion likely
dominates the high-column density \ion{H}{i} absorbers at $z=0.25$.

In summary, we see distinct trends in the type of material traced by
absorption as a function of ionisation potential.  Low metal ions,
particularly close to galaxies, generally arise from recycled accretion,
while high ion absorption is from ancient outflows.  This latter finding
is consistent with the analysis of \cite{tum11}, who detect large
amounts of \ion{O}{vi} in the CGM and determine that it must have been generated
by protracted earlier epochs of star formation and winds.

\subsection{Fractional Column Density Distributions}

\begin{figure*}
 \includegraphics[width=5.4in]{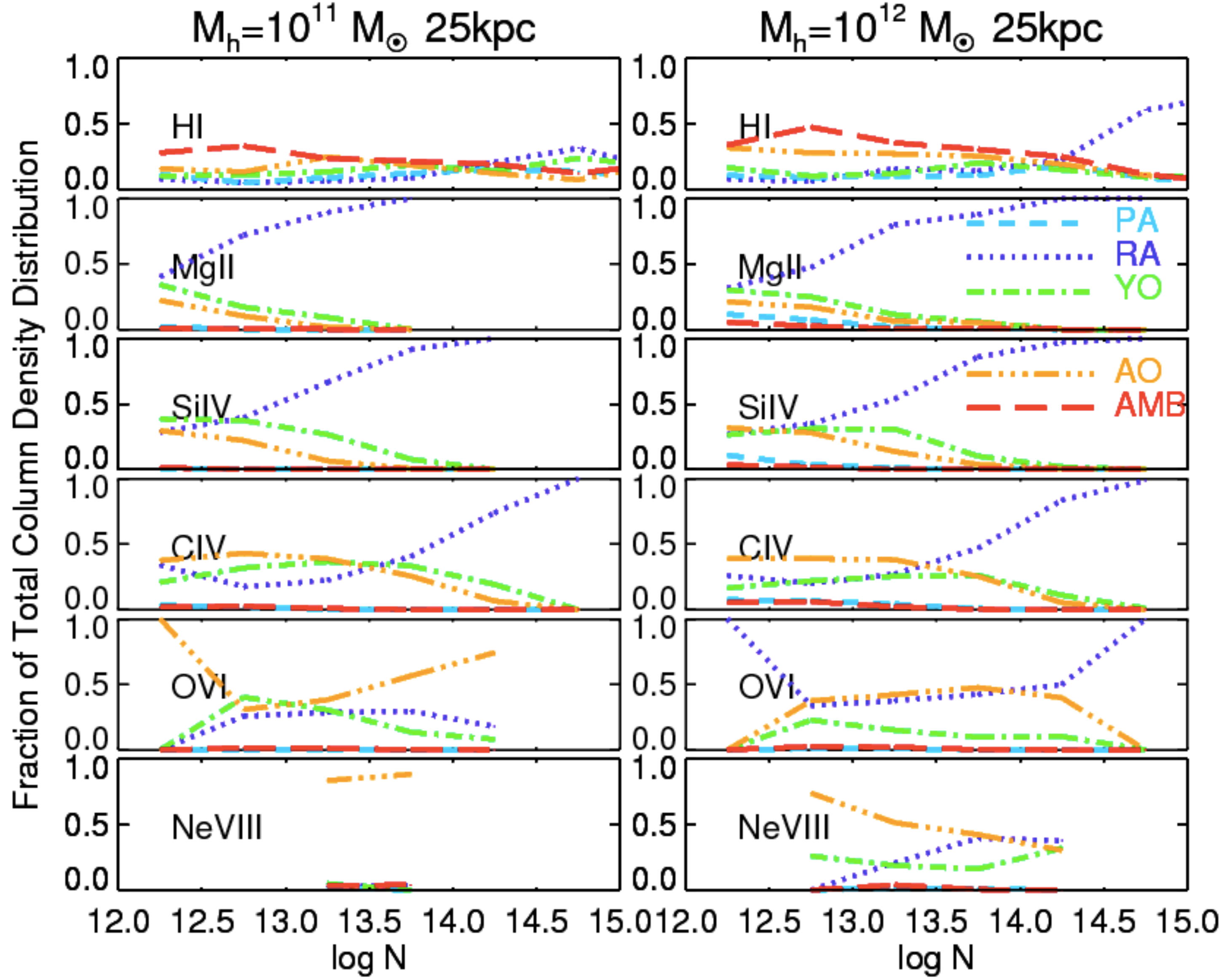}
  \includegraphics[width=5.4in]{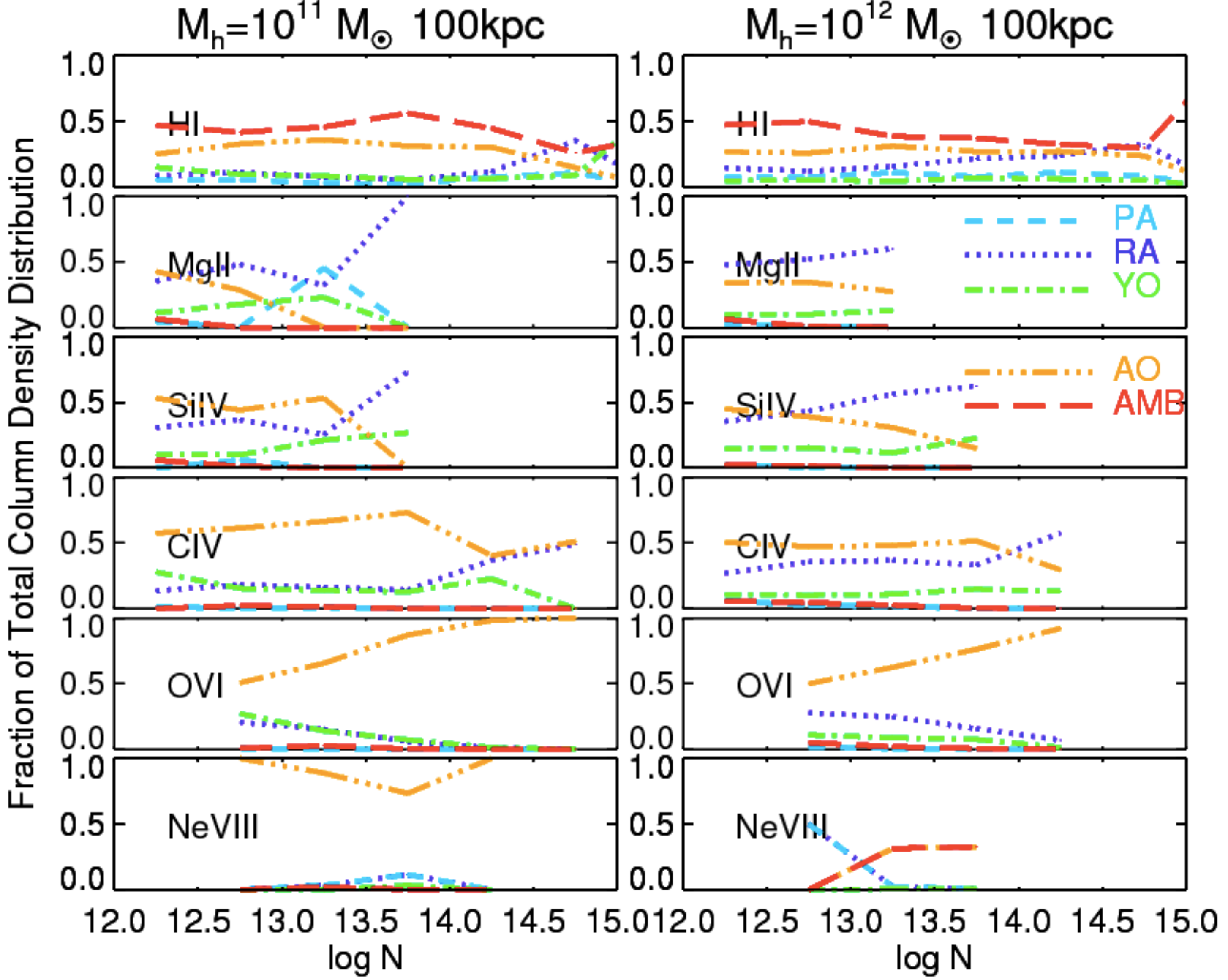}
\caption{Fractional column density distributions, i.e. the fraction of 
absorption in bins of size $\log N=0.5$ owing to the various categories for
different species (as labelled)
at impact parameters of 25~kpc (top panels) and 100~kpc (bottom panels), for \mlo~halos 
(left panels) and \mhi~halos (right panels). Where lines do not extend, there is no column density in that bin from any gas.}
\label{cdd}
\end{figure*}

Column density distributions (CDDs) provide a more detailed look
at the strengths of individual absorbers within our various categories.
Figure \ref{cdd} shows the ``fractional CDD" for each of our
ions, for our two halo mass bins (\mlo~left panels, \mhi~right panels), at an
impact parameter of 25~kpc (upper panels) and 100~kpc (lower panels).
The lines are colour-coded by category: pristine accretion (light blue), 
recycled accretion (dark blue), young outflows  (green),
ancient outflows (orange), and ambient gas (red).
We define fractional CDD as the fraction
of the total absorption {\it in that column density bin} 
that owes to a given category. Where colored lines do not extend, there are no absorbers in that particular column density bin. In some cases the lines do not add up to one, that is because occasionally there is contribution from ISM gas (not plotted here) to the total. For \ion{Ne}{viii} at 100~kpc, we note that the colored lines overlap in some cases. 

In these plots we can identify some general trends as a function
of ionisation potential.  For \HI, there is more strong absorption from recycled accretion than pristine accretion. At impact parameters of 25~kpc in \mhi~halos
the strongest metal lines are almost always dominated
by recycled accretion.  The trend is particularly strong for low
ions and is even true for \ion{O}{vi}, although strong \ion{Ne}{viii} lines
have comparable contributions from both young and old outflows.
Strong metal absorption at this impact parameter
traces material that will fall into the galaxy within a few Gyr;
just as for the strongest \ion{H}{i} absorbers.
At lower metal columns outflows become important and can even dominate.
Generally young outflows are more important for lower ions and ancient
outflows are more important for higher ions. Low column HI absorption owes
mostly to ambient gas and less to ancient outflows.
These trends also hold for \mlo~halos except that ancient outflows now
dominate at all columns for the high ions.

At an impact parameter of 100~kpc for \mhi~halos,
the trends look somewhat different.
For low ions the strongest observable metal lines
still come from recycled accretion, but the contribution from outflows can be
significant at lower columns, and at this larger impact parameter it is almost
always ancient outflows that are more important than young outflows.
It is likely that the low ions still arise from gas close to galaxies,
but perhaps close to substructures such as what is seen in Figure~\ref{pretty}.
In contrast, \ion{O}{vi} is completely dominated by ancient outflows, 
consistent with the findings of \cite{opp09}.
Young outflows contribute but do not dominate at any column density.
For our mid ion \ion{C}{iv}, the main contribution is from ancient outflows except for very strong absorbers around \mhi~halos, where recycled accretion is prominent. Meanwhile,
the \ion{H}{i} is usually dominated by ambient gas with important
contributions from ancient outflows for both halo masses with recycled
accretion also becoming important at higher columns.

\subsection{Covering Fractions}

A key observable is the covering fraction of absorbers down to some
particular column density or equivalent width limit.  Here we sum
all absorbers within $\pm$ 300 km/sec of the targeted galaxy
for each of our lines of sight, above an equivalent width limit 
of 0.05\AA. We use summed equivalent widths rather than individual components to minimize sensitivity to details of the line identification and deblending algorithm.

\begin{figure*} 
 \includegraphics[width=7in]{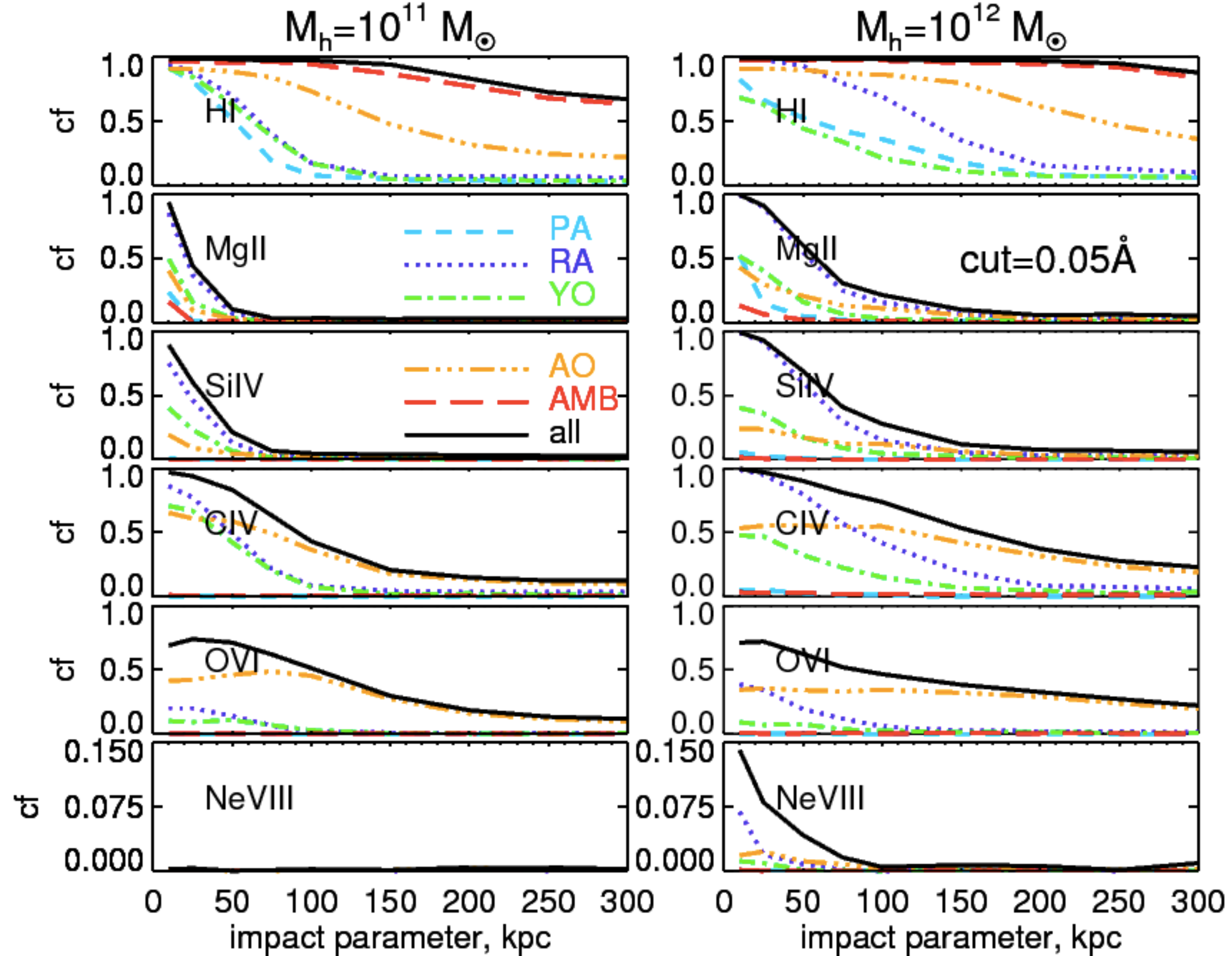}
\caption{Covering fractions for \mlo~(left panels) and \mhi~(right panels)
 halos, showing 
the fraction of sight lines with equivalent widths $> 0.05$\AA\
for
the species as labelled. The black line includes all the gas
and the broken coloured lines are split into the categories as 
labelled. Note the different y-axis range for \ion{Ne}{viii}, and note that individual categories are computed separately and do not sum to the black line.}
\label{cf}
\end{figure*}

Figure \ref{cf} shows the covering fractions for our various
ions, for our two halo mass bins. The total is shown as the black line and can be compared to what is observed. The 
coloured lines show the theoretical covering fraction including only the gas in the labelled category. Note that the coloured lines do not sum to the black line -- we compute the
covering fraction of each category individually above 
our chosen equivalent width limit of 0.05\AA.

\HI\ has a near-unity covering fraction at all impact parameters
and halo masses, owing mostly to ambient gas. Ancient outflows have large covering fractions within $\sim 100$kpc for
both halo masses and recycled accretion also becomes substantial for \mhi~halos.  Pristine accretion and young outflows also have comparable non-trivial
covering fractions for both halo masses but it
is clearly much smaller than recycled accretion in \mhi~halos.  Hence even 
\HI\ does not typically trace pristine inflows in these halos, because as we
showed in Fig~\ref{pie}, the accreting mass budget is actually dominated
by recycled accretion at these epochs.

As expected, the metal lines almost never show significant covering fractions
from either ambient or pristine accretion, as the metal content tends to
be quite low.  The remaining categories follow trends similar to
those seen for the column densities:  low ionisation line covering
fractions tend to be almost always dominated by recycled accretion, 
while \ion{C}{iv} is dominated by recycled accretion close in but by ancient outflows
further out.  \ion{O}{vi} is dominated by ancient outflows everywhere, except for at very low impact parameters around \mhi~ halos .  Qualitatively,
this trend holds for all halo masses, although the overall metal line
covering fraction, particularly at small impact parameters, falls 
towards smaller halos.

These results are intended to qualitatively illustrate the behaviour
of covering fractions, and show that they mimic trends seen in other
statistics.  We will present a more detailed comparison of covering
fractions versus COS-Halos data, including covering fractions as a function of
equivalent width, using simulated spectra chosen to match the COS-Halos
sample, in an upcoming paper (Ford et al. 2014, in preparation).

\section{Physical Conditions}

\subsection{Mass and metal profiles}

\begin{figure*}
 \includegraphics[width=7in]{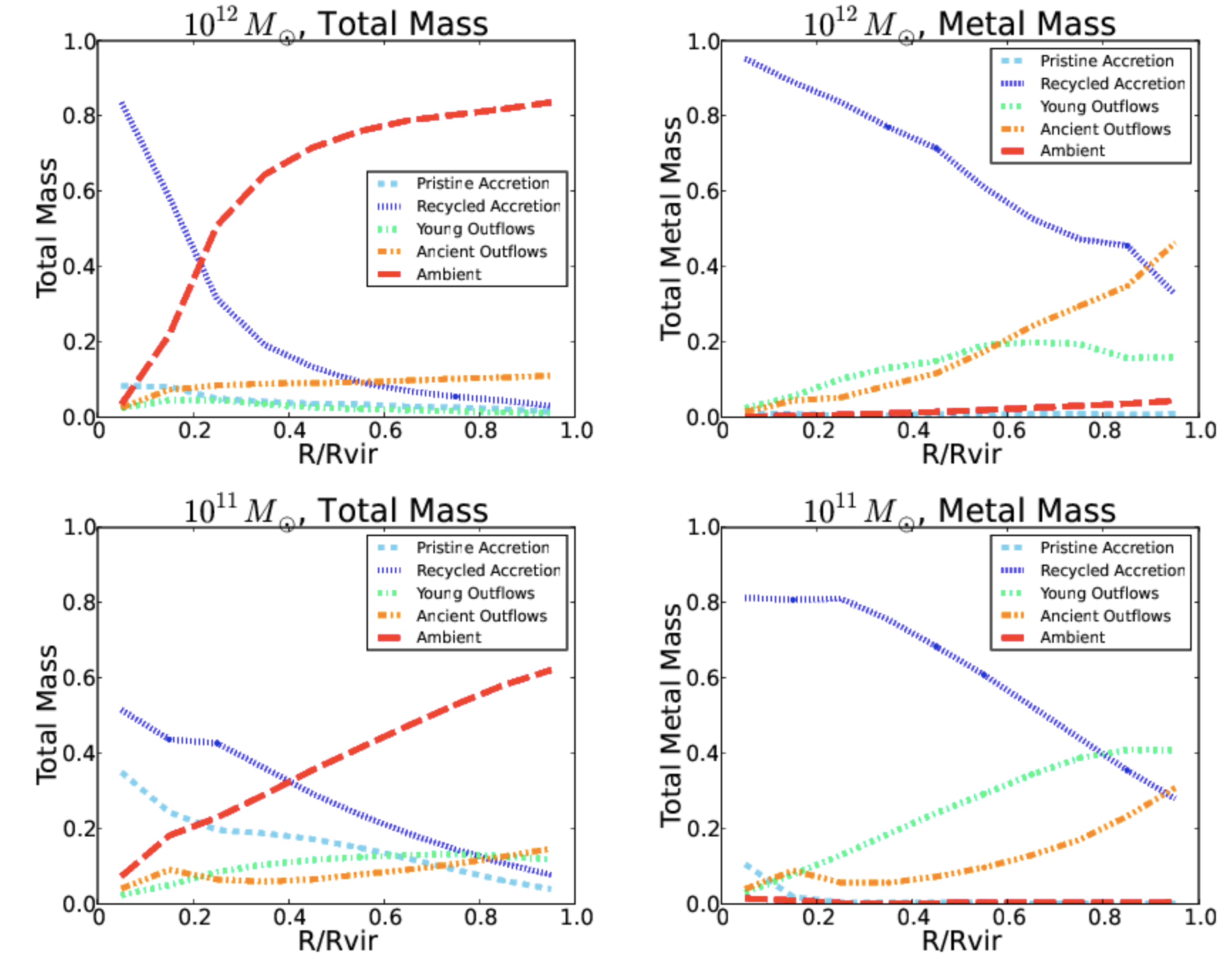}
\caption{ Left panels: fraction of the total mass of non-ISM gas particles in various 
categories as a function of $R/R_{vir}$ at $z=0.25$, as labelled. Right 
panels: fraction of the total {\it metal} mass. Top panels are for \mhi~halos and the
lower panels are for \mlo~halos. Only central galaxies are considered.}
\label{rmask}
\end{figure*}

To better understand why the absorption patterns detailed in the previous
section arise, we now examine the physical conditions in the halo
associated with the inflowing and outflowing gas.  We begin by considering
the mass and metal profiles associated with our various categories.
Figure \ref{rmask} shows these profiles as a function of $R/R_{\rm vir}$, 
out to the virial radius. We emphasize this is a 3-D radial profile, not projected as impact parameter. The left panels show the mass fraction as a
function of radius in each category, while the right panels show the
metal fractions.  The two rows show our two halo mass bins. We include only gas around central galaxies, since our LOS are only around central galaxies. As explained in \S 3.3, each gas particle is assigned to a galaxy to which it is most bound. If a particle is assigned to a satellite galaxy we do not include it in Figure \ref{rmask}.

Let us first consider recycled accretion.  Broadly speaking, for both mass
bins, the percent of gas mass and metal mass in recycled accretion
decreases with increasing $R/R_{\rm vir}$.  This soon-to-be
recycled material is
fairly close to galaxies, which explains why it is predominantly
found in cooler, denser gas and explains why \HI\ at low impact parameters
comes more from recycled accretion than from ambient gas.
In Figure \ref{NvsB} we noted that,
at least for the more massive halos, metal absorption from recycled
accretion is fairly strong close to galaxies, while ancient outflows
become more dominant at larger impact parameters. This is why in
the case of the \mhi~halo at low impact parameters there is substantial absorption
in \ion{O}{vi} from recycled wind 
material -- because very close to \mhi~galaxies,
most of the metals are found in recycled accretion material.
Hence, even though much
of the recycled wind material close to galaxies is cool and dense,
the high enrichment level still allows for significant 
\ion{O}{vi} absorption.  Meanwhile, the low metal
ions have rapidly dropping absorption profiles with radius, while high ions have
radial profiles that are
more flat.  The low metal ions necessarily arise in cool, enriched gas, which typically has a short recycling time. In \mhi~halos the recycled accretion is peaked within r$<$0.3${R}_{\rm vir}$, while for \mlo~it is spread more smoothly out to ${R}_{\rm vir}$.

For pristine accretion, we see that the percentage of the gas and metal 
mass also drops with increasing $R/R_{\rm vir}$. 
The percentage of mass in ambient gas increases with radius, becoming the
dominant component at large radii, but the ambient gas contains few
metals.
The percentage of the gas mass in ancient outflows changes little with radius,
but the percentage of the metal mass increases appreciably.
The percentage of mass in young outflows is fairly flat with radius but the
metal mass rises.  These profiles set the baseline for where metals are
located within the CGM, which is then convolved with the physical 
conditions to give rise to the actual amount of absorption.

\subsection{Velocity and kinematics}

\begin{figure*} 
 \includegraphics[width=5.5in]{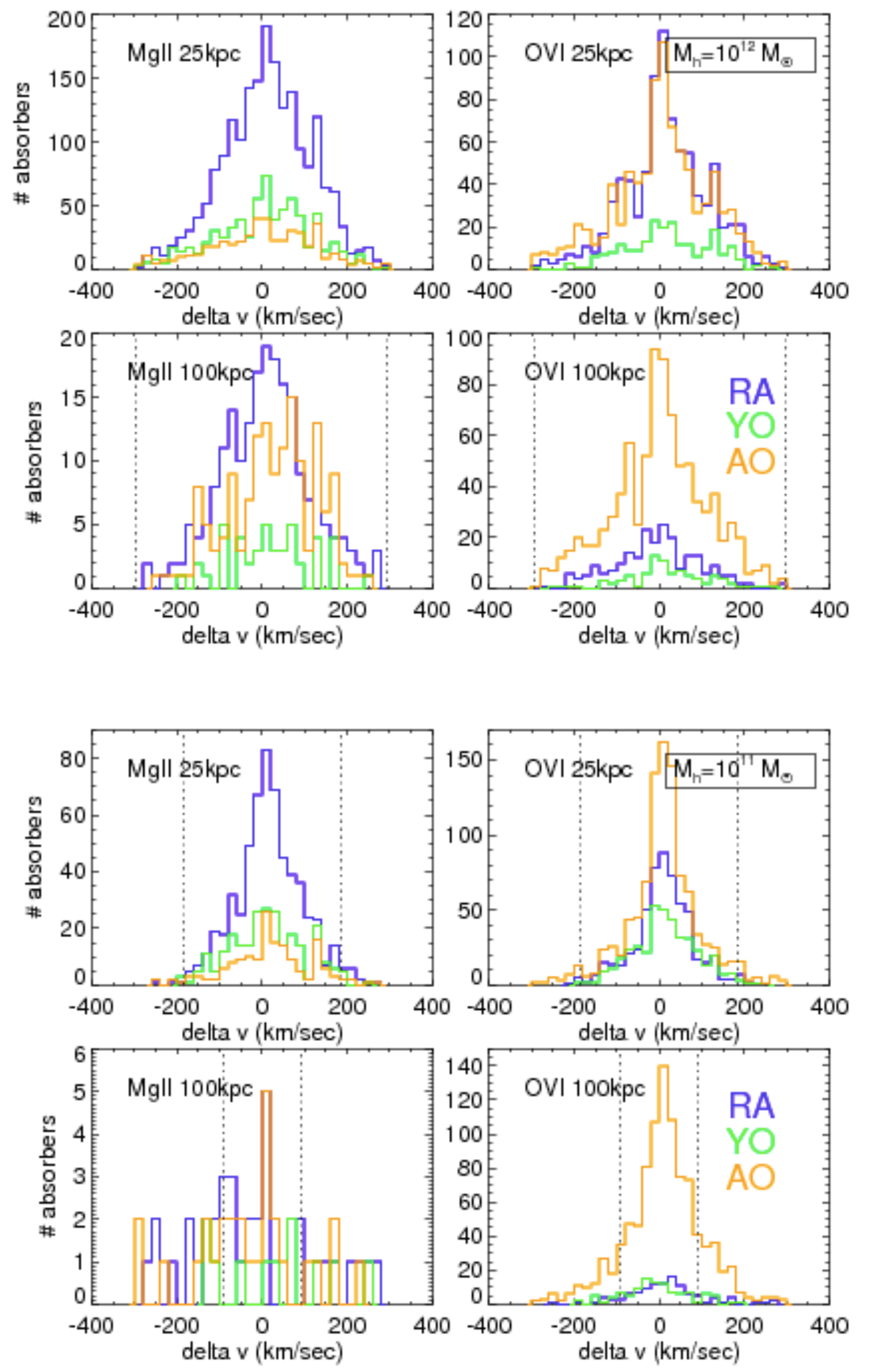}
  \caption{Histograms of velocity, relative to the galaxy systemic velocity, of absorbers split into recycled accretion (blue), young outflows (green), and ancient
outflows (orange), for \ion{Mg}{ii} \& 
\ion{O}{vi}, at 25~kpc \& 100~kpc as labelled for \mhi (top panels) and \mlo
(bottom panels) halos. Vertical 
lines demarcate the escape velocity at the midpoint of each mass bin for R=25~kpc or 100~kpc as labelled. (Escape velocity for \mhi~halos at 25~kpc, 590 $\kms$, is outside the range of the plot). Note the 
different y-axis range for each panel.}
 \label{velhist}
 \end{figure*}

We can also track the kinematics of our absorbers, as shown in Figure 
\ref{velhist}. In this Figure we focus on \ion{Mg}{ii} and \ion{O}{vi}, as 
examples of low and high ions. We include only recycled accretion (blue), young 
outflows (green), and ancient outflows (orange), since the other categories do 
not contribute significantly to the metal absorption. To measure the
kinematics of CGM absorbers relative to their galaxy,
we subtract 
the galaxy's systemic velocity and plot a histogram of the velocities of the 
absorbers. Here we show components, not systems, and all absorbers above 30 m\AA\ are 
plotted. We show both halo bins, \mlo and \mhi, at impact parameters of
25~kpc and 100~kpc. For comparison we show the approximate escape
velocity ($v_{esc}=\sqrt{2GM/r}$) as dashed lines, using the impact parameter (25~kpc or 100~kpc) as the radius and the midpoint of each mass bin as the mass. 

For both the \mhi~and \mlo~halos we see that most of our absorbers lie within 
the escape velocities of their host halos, at least at these impact parameters,
for both \ion{Mg}{ii} and \ion{O}{vi}. We note that for \mhi~halos, this is true by definition, as we only consider those absorbers within $\pm$ 300 $\kms$ as associated with a galaxy, and the escape velocity at these impact parameters is $\ge$ 300 $\kms$. However, one can see the distribution trailing off; even if we did not make this cut at 300 $\kms$ it is unlikely we would see significant numbers of absorbers past 300 $\kms$. In general, the shape of the velocity distributions
for the different categories are similar, showing that there is no strong
kinematic trend that distinguishes inflow from young or ancient outflow.

\subsection{Phase Space Plots}

\begin{figure*}
\includegraphics[width=6.3in]{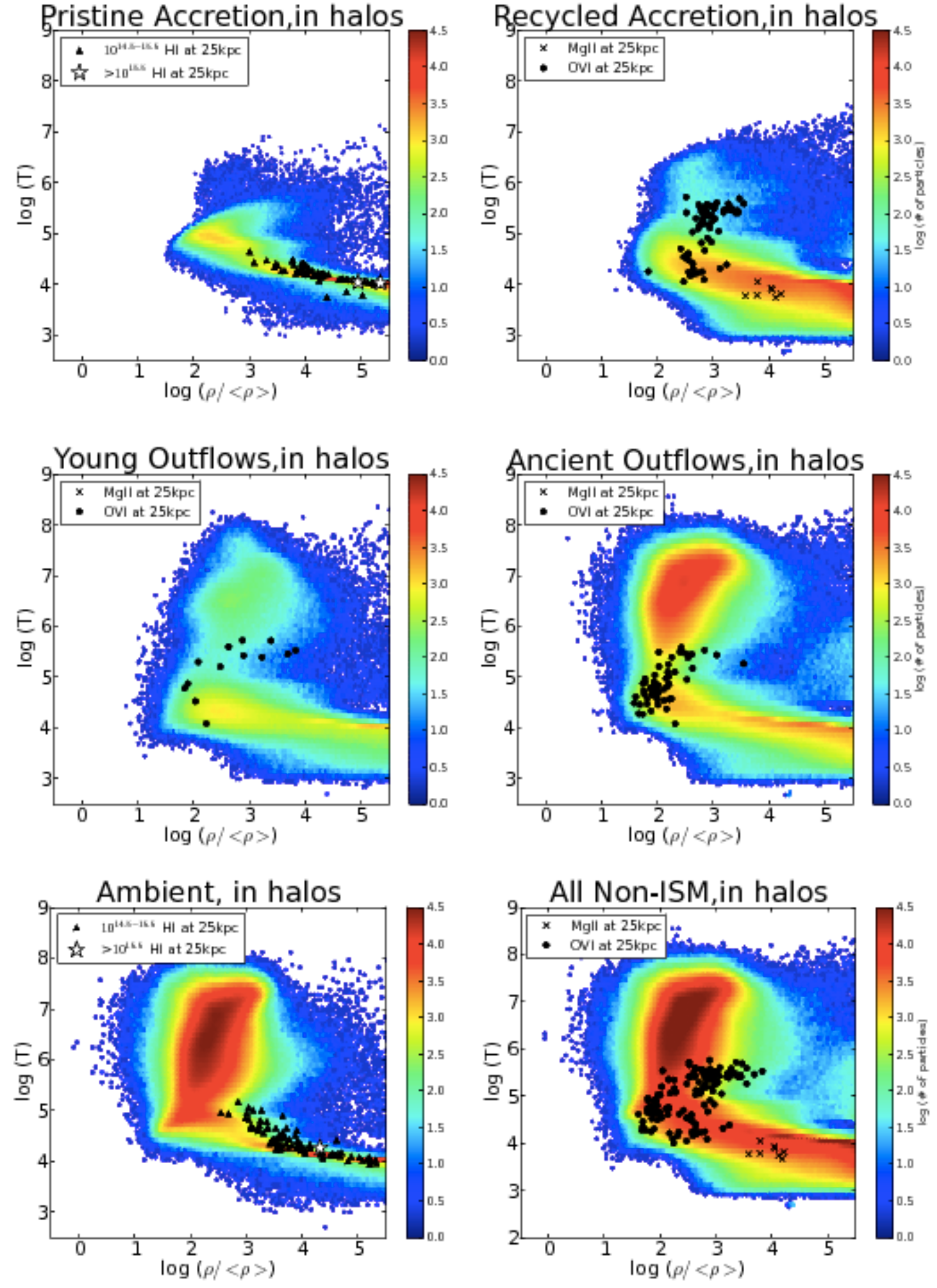}
\caption{Distribution of gas in temperature-density phase space, for our five categories and for all non-ISM gas, as labelled. We overplot the location of strong (N $>$ ${10}^{14}$) absorbers of \ion{Mg}{ii} and \ion{O}{vi} in \mhi~halos at 25~kpc, for all non-ISM gas, as well as broken out by category for recycled accretion, young outflows, and ancient outflows. In the pristine accretion and ambient panels, where there is no strong \ion{Mg}{ii} or \ion{O}{vi} absorption in \mhi~halos at 25~kpc, we overplot the location of \ion{H}{i} absorbers of various strengths as labeled. The colourbar is the same in each panel, showing the log number of particles.}
\label{phase25}
\end{figure*}

\begin{figure*}
\includegraphics[width=6.3in]{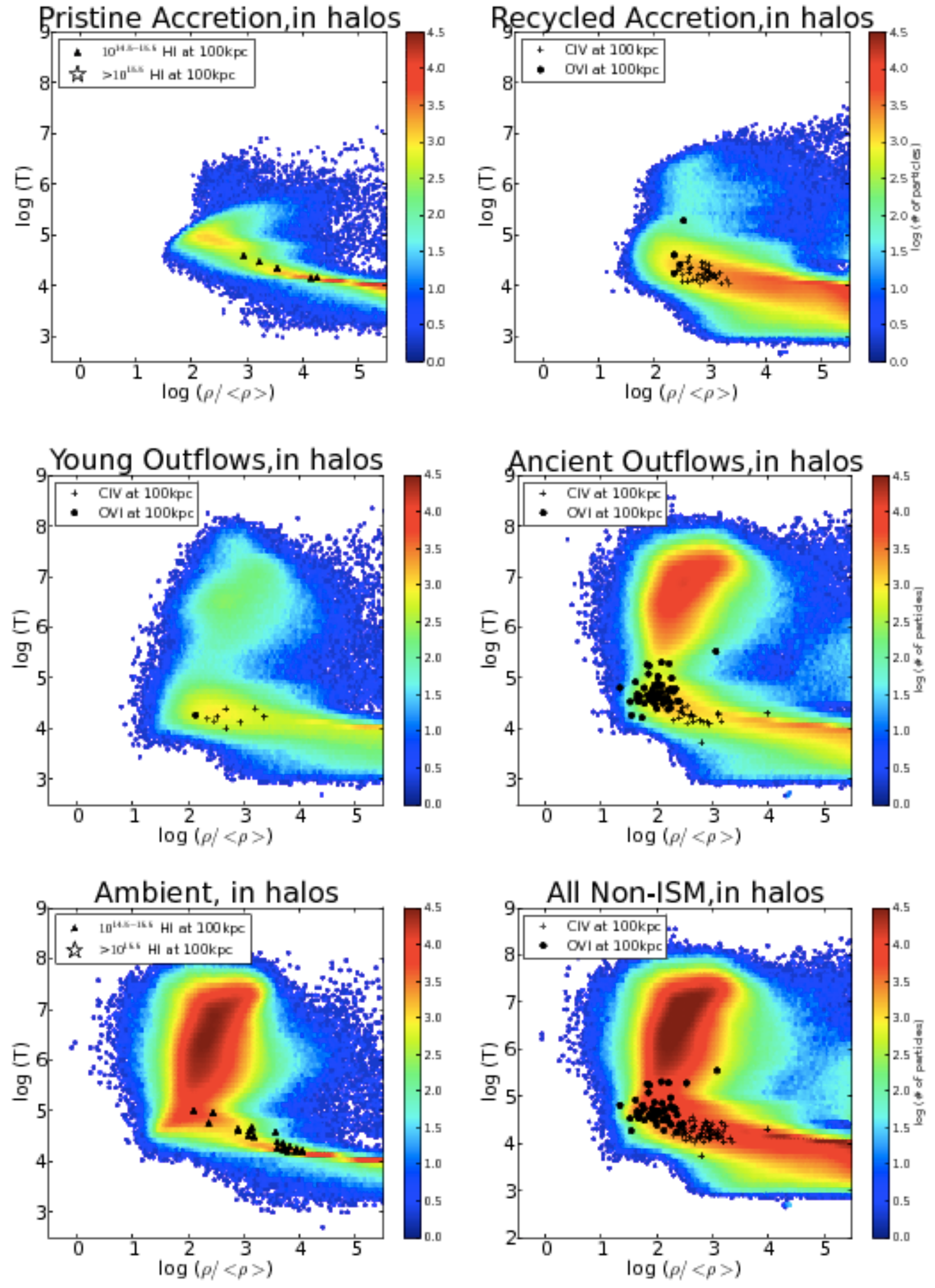}
\caption{Phase space distributions as in Figure \ref{phase25}. We overplot the location of strong (N $>$ ${\rm 10}^{14}$) absorbers of \ion{C}{iv} and \ion{O}{vi} in \mhi~ halos at 100~kpc, for all non-ISM gas, as well as broken out by category for recycled accretion, young outflows, and ancient outflows. In the pristine accretion and ambient panels, where there is no strong \ion{C}{iv} or \ion{O}{vi} absorption in \mhi~halos at 100~kpc, we overplot the location of \ion{H}{i} absorbers of various strengths as labeled. The colourbar is the same in each panel, showing the log 
number of particles.}
\label{phase100}
\end{figure*} 

A key diagnostic of the physical conditions of an absorber is its
location in the temperature-overdensity phase space.  In \citet{for13} we extensively examined the
location of absorbers in this phase space using similar simulations.  Unsurprisingly, we
found that low ionisation potential absorption like \ion{Mg}{ii} arises in
cooler, denser environments, while high ionisation potential absorption
such as \ion{O}{vi} and
\ion{Ne}{viii} exist in warmer, more diffuse environments,
with \ion{Si}{iv} and \ion{C}{iv} being intermediate~\citep[see Figure 6 of][]{for13}.  Here we expand on that work by examining where our
different inflow/outflow categories lie in phase space, and which
absorbers might be tracing such gas.  We focus here on \ion{Mg}{ii},
\ion{C}{iv}, and \ion{O}{vi} as representative of low, mid, and
high ionisation potential lines; other transitions follow trends based on
their ionisation potential.  We will also examine \HI\ in pristine
accretion and ambient gas, since this gas generally gives rise to
weak if any metal absorption.

In Figure \ref{phase25}, we show phase space contours of halo gas in each category at $z=0.25$ (first 5 panels), as well as all non-ISM gas particles
(last panel) within halos.  The legend for the
coloured contours is indicated on the right, showing the logarithmic
number density of particles at each position in phase space; for ease of 
comparison it is the same in all panels, even though not all the categories
have the same total number of particles. 

In the recycled accretion, young outflow, and ancient outflow panels
we overplot, as the black symbols, the
location in density and temperature space of \ion{Mg}{ii} (x's),
and \ion{O}{vi} (filled circles) absorbers
that arise from particles in that category.  For clarity, we
have only plotted the strong absorbers (N $>$ ${\rm
10}^{14}$ ${\rm cm}^{-2}$), and only around \mhi~halos at impact
parameters of 25~kpc. In the pristine accretion and ambient panels, there are no strong \ion{Mg}{ii} or \ion{O}{vi} absorbers around \mhi~halos at 25~kpc. Instead, we plot the \ion{H}{i} absorbers around \mhi~halos at 25~kpc, for the strengths as labelled. 

Both pristine (upper left) and recycled (upper right) accretion consist almost exclusively
of cooler ($T<10^5$~K) gas.  Gas at temperatures $T>{10}^{5}$~K  is unlikely to accrete onto
galaxies between $z=0.25$ and $z=0$.  Nonetheless, there is some recycled accretion
gas extending up to hotter temperatures at moderate overdensities
($\sim 10^3$), where the cooling time becomes sufficiently short
that gas can radiate away its thermal energy. 

Considering metal lines, \ion{Mg}{ii} absorbers trace the densest recycled accretion gas,
typically with $\delta>10^4$, while \ion{O}{vi} traces more diffuse and often significantly hotter halo gas. Both \ion{O}{vi} and \ion{Mg}{ii} arise from recycled accretion at 25~kpc, but never from the same phases of this gas! 
Much of the absorption arises in cooler photo-ionised gas, but the hotter plume
of recycled accretion gas mentioned above gives rise to substantial \ion{O}{vi}
absorption as it transitions through the collisional ionisation
fraction maximum for \ion{O}{vi} at around 300,000~K.  Therefore, we
expect a significant amount of collisionally-ionised oxygen arising
in halos, though as we argued in \citet{opp12} and \citet{for13}
it is not globally dominant relative to photo-ionised \ion{O}{vi} absorption. We emphasize that we have only plotted the strong absorbers here, which for \ion{O}{vi} is often collisionally ionised. At 25~kpc there is substantial 
strong \ion{O}{vi} absorption from both recycled accretion and ancient 
outflows, while at larger impact parameters, it chiefly arises
in ancient outflows, (Figure \ref{NvsB} and in Figure 
\ref{phase100} below). 

Young and ancient outflows (middle panels) are generally at somewhat lower
densities, and
they show more significant amounts of gas near the virial temperature.  Note
that in lower mass halos (not shown as a separate subdivision), there is very little of this
hotter gas component~\citep[e.g.][]{ker09}. \ion{O}{vi} absorption
still comes from the same general region in these panels as in
the recycled accretion panel: warm, low density gas that is both photo-ionised 
and collisionally-ionised. Although ancient outflows do have some
material roughly in the right density/temperature range for strong \ion{Mg}{ii}
absorption, we note 
that close to galaxies there is very little metal mass in ancient outflows,
as shown in Figure \ref{rmask}. 

In the lower panels we see even hotter material in the
ambient gas (left panel). This material generally has too low a metal 
content to provide much strong \ion{O}{vi} absorption and, furthermore, it has
been shock heated to temperatures that are too high for strong \ion{Mg}{ii}
absorption.  It is interesting that most of the
\ion{O}{vi} absorption in our models comes from infalling or (formerly)
outflowing gas, not from ambient halo gas.  In part this owes to
our definition of ambient gas as never having been in 
an outflow, so it is very difficult for this gas to become enriched.
Nonetheless, this indicates that there may be a substantial amount
of warm-hot gas within halos that is not traceable via metal
lines, and one must rely on, e.g., broad \HI\ absorbers to 
characterise this gas \citep[e.g.][]{ric12}. 

In Figure \ref{phase100} we show the same coloured regions as in
Figure \ref{phase25}, but have overplotted absorption at impact parameters
of 100~kpc instead of at 25~kpc; at larger impact parameters, ancient
outflows dominate the absorption in high ions. Because there is
almost no \ion{Mg}{ii} absorption at 100~kpc, we choose instead to
plot \ion{C}{iv} with \ion{O}{vi}. One can see that \ion{O}{vi} comes from warmer, less dense gas, but is rarely collisionally ionized at temperatures $>$ ${10}^{5}$~ K.  \ion{C}{iv} absorption resides at intermediate overdensities between \ion{O}{vi} and \ion{Mg}{ii} as shown in the previous figure. 

These plots help answer the question: why is \ion{O}{vi} absorption
mostly coming from ancient outflows? It is because ancient outflows
have more material at the temperatures and densities favourable for
\ion{O}{vi} formation (and for the formation of high metal ions in
general).  Conversely, absorption in the low ions largely comes
from cold, dense gas that is likely to accrete, so it arises in the
recycled accretion category rather than the outflow category.
Intermediate ions can have significant contributions from recycled
accretion, young outflows, or ancient outflows, depending on the
impact parameter.

Moving to \HI\, we see that at 25~kpc (Fig \ref{phase25}) the strong
($N > 10^{14.5}\cdunits$) \ion{H}{i} absorbers for pristine accretion arise
at high densities ($\delta \ge 10^3$). In the relation between
overdensity and \HI\ column density given by \citet{dav10}, \HI\
absorbers with $N={10}^{14.5-15.5}~{cm}^{-2}$ are predicted to lie at 
$\delta=10^2-10^3$, albeit with substantial scatter.
Why does pristine accretion not have absorbers in this
overdensity range? The absorbers clearly follow the overall peak
of the pristine accretion distribution (red/yellow region), and
there is simply not enough pristine accretion material in the
that overdensity regime to give rise to \HI\ absorption.

This is different from their distribution in the ambient case, where
there is strong \HI\ at not only high overdensities, but also at
intermediate overdensities of
$\delta= 10^2-10^3$. This is because there is more mass in
ambient material than in pristine accretion material, as shown in
Figure \ref{rmask}. The strong absorbers in ambient material can
lie at $T>$ ${10}^{4.5}$~K, and there is more spread in the
temperatures as well: pristine accretion absorbers are all at
temperatures
$T<10^{5}$~K, while some ambient absorbers arise in hotter gas $T>10^{5}$~K,
though most are from cooler gas. In both pristine accretion and
ambient, the very strong ($N>10^{15.5}\cdunits$) absorbers (two for pristine
accretion, one for ambient) exist at high overdensities,
$>10^{4}$, consistent with the relation given by \citet{dav10}:
stronger \HI\ absorption comes from highest density gas, close to
galaxies.  Note that if we were to display \HI\ absorbers in the 
recycled accretion panel, there would be significantly more such
strong absorbers at these overdensities.

By 100~kpc (Fig. \ref{phase100}), there is much less strong \HI\,
and no very strong \HI\, in either category. This is because the
overdensity drops with radius, so at larger impact parameter there
is less dense material to give rise to \HI\ absorption. There are
more absorbers in ambient than in pristine accretion, because at
large radius there is more ambient than pristine material, as seen
in Figure \ref{rmask}. Like the low/mid metals, \HI\ absorbers
generally follow the peak of the gas distribution, excluding areas
that are too hot.

Overall, a given metal ion will absorb at similar densities and
temperatures across all categories.  Whether there is substantial
absorption in a given ion from a given category largely depends on
how much material in that category lies at the right temperature/density
combination for absorption.  It is also a function of the total
amount of metal mass in a category at a given radius from the galaxy,
as discussed in the previous section.  Hence in principle, a suite
of 
absorbers spanning a range of ionisation states can probe the full range 
of physical conditions as a function
of radius in the CGM, with the exception of hot unenriched gas.

\section{Numerical Considerations}

Our predictions depend on both our physical model of galaxy formation
and our numerical implementation of that physical model.
Key elements of the physical model include
the $\Lambda$CDM cosmological framework, the adopted cosmological
parameter values, standard hydrodynamic and radiative cooling processes, the
star formation prescription, and, critically, the ezw wind prescription
that sets mass loading factors and ejection speeds.  In the absence
of winds, different implementations of SPH, adaptive mesh refinement
(AMR), and moving mesh hydro simulations show reasonable but not perfect
convergence on the mechanisms and rates of galaxy growth.
All methods show that halos with $M \la 10^{11.5} M_\odot$ are fed
by filaments of cold ($T \sim 10^4\,$K) gas that penetrate far
inside the halo virial radius, and that large halos of shock heated
gas (typically $T \ga 10^6\,$K) develop only at higher halo masses
(e.g., \citealt{ker05,ker09,ocv08,age09,bro09,cev10,
nel13}).  This cold-to-hot transition can be understood analytically
in terms of the ratio of gas cooling time to the halo dynamical time
\citep{bin77,ree77,sil77,whi78,whi91,bir03,dek06}.
In the critical regime near $10^{12} M_\odot$, there is some disagreement among numerical methods on whether cold filaments persist
all the way to the central galaxy \citep{nel13}, though this appears
to be a shifting of transition boundaries in mass and redshift rather
than a universal difference of behaviour.\footnote{For example, the
$10^{11.5} M_\odot$ shown by \cite{nel13} at $z=2$ (their figure 7)
looks significantly different between GADGET (SPH) and AREPO (moving mesh),
but the $z=3$ snapshots of the same halo are nearly identical between
the two codes (D.\ Keres 2013, presentation at IAP Colloquium on Origin of the Hubble Sequence).}
Another significant
difference, probably more important for galaxy growth rates, arises in
the density profiles and consequent cooling rates of the shock heated
gas component.  The entropy-based formulation of SPH \citep{spr02}
implemented in GADGET-2, and thus in our simulation here, appears
to be especially effective at maintaining separation of fluids in
different phases and reducing cooling of the hot gas component,
relative to other formulations of SPH or to the moving mesh code AREPO \citep{ker09,nel13}.

While numerical differences of this sort would have some quantitative
impact on the kinds of predictions presented here, the more important
numerical uncertainties have to do with the implementation of galactic
winds themselves.  As discussed in \S 2, we follow \cite{spr03}
in implementing winds via ejection of particles from the star-forming
ISM.  Because these wind particles tend to be metal-rich, and because
they leave the galaxy at ISM temperatures, they typically have
short cooling times and remain cold even when they interact
with a surrounding hot halo.  Wind particles retain their heavy elements
by construction, so there is mixing with the ambient halo gas only
to the extent that the particles representing these components become mixed.  This behaviour may be a reasonable representation of
reality, but it is also possible that instabilities mix wind gas
and ambient halo gas on scales below the resolution of our simulation
(and perhaps below the resolution of any current cosmological
simulations of galaxy formation).

There are strong circumstantial arguments for galactic winds with
phenomenology roughly like that of our ezw model, including the
low observed ratios of stellar mass to halo baryons, the widespread presence of intergalactic and circumgalactic metals, and direct observations and
high-resolution simulations of galaxy outflows.  We therefore expect that the global
mass and metal budgets in our simulation are approximately correct.
However, the detailed density and temperature structure of ejected
and recycling gas, and therefore the absorption in individual ionic
species, is necessarily more uncertain, and our predictions here should
be regarded as specific to our physical model and numerical implementation
of winds.  In future work we will use a wind formulation that incorporates
an explicit subgrid model for mixing of metals and exchange of thermal
energy between wind particles and their neighbours, which will enable
us to investigate a range of possible behaviours.  Extensive mixing
would blur the distinctions between ``outflow'' gas and ``ambient'' gas,
and between ``recycled'' and ``pristine'' accretion, complicating
some of the descriptions we have adopted in this paper.
More generally, numerical simulations of stellar feedback and the
baryon cycle are in a phase of rapid progress, with an increase in
the sophistication of models and the range of numerical approaches.
We expect that comparison of results from different simulations,
and comparison of these results to rapidly
improving observations of circumgalactic gas and galactic outflows,
will lead to numerous insights and much progress over the next several years.

\section{Conclusions}

We examine the physical and dynamical state of gas around galaxies
in a $32\hmpc$, quarter-billion particle cosmological hydrodynamic
simulation.  We concentrate on $z=0.25$, and use past and future
simulation snapshots to delineate which particles have been ejected
from galaxies (and how long ago), which particles are falling into
galaxies by $z=0$, whether or not they have previously been part
of a galaxy's ISM, and which particles are ambient material not
participating in the baryon cycle.  We use these categories to
examine both observable and physical properties of the gas, to
elucidate a connection between observational absorption line probes
of the CGM and its dynamical state, i.e. whether it is inflowing,
outflowing, or ambient. As in our earlier simulations \citet{opp10},
recycled accretion of enriched gas that is ejected in winds and
then re-accreted plays a major role in low-redshift galaxy accretion.

Our main findings are as follows:

\begin{enumerate}

\item Most of the absorption of low-ionisation potential metal
species owes to enriched material that will fall back into galaxies
within a few Gyr. This occurs because low-ionisation species
preferentially trace cold, dense, enriched environments, and
hydrodynamic interactions generally cannot prevent this gas from
accreting.

\item Most of the absorption of high-ionisation potential metal
species owes to outflows ejected much earlier than the present
epoch, ``ancient outflows". This occurs because high-ionisation
potential species tend to be prominent in warmer, more diffuse
environments. Some of this high-ionization material is also
re-accreting onto galaxies.

\item The metal mass fraction in recycled accretion drops off steeply
with $R/R_{vir}$ (see Figure \ref{rmask}), which gives rise to a
rapid drop in absorption with impact parameter for low ions. The
metal mass in ancient outflows increases mildly with $R/R_{vir}$,
which partly explains why the absorption profile of the high ions
stays relatively flat out to the virial radius.

\item Low-mass halos ($<$${10}^{11.5}$ $\msolar$) have a greater
proportion of recycled material than high-mass halos ($>$${10}^{11.5}$
$\msolar$). Even though high-mass halos assemble proportionally
more of their mass via recycled accretion than pristine accretion
\citep{opp10}, recycled accretion represents a proportionally larger
fraction of the circumgalactic medium around low-mass halos.

\item Where absorption of a given ion is found, that absorption
exists at roughly similar densities and temperatures among the
various categories.  Hence, absorbers with a range of ionisation
potentials can broadly trace out the density and temperature structure
of the CGM.  The strengths of various absorption lines within various
inflow/outflow categories primarily reflects the phase space location
and metallicity of the gas in those categories.

\item~\HI\ predominantly traces non-accreting material that has
never been in a wind, i.e. ``ambient".  Ambient material holds the
majority of mass in the halo and is quasi-spherical, which is why
it dominates the \HI\ absorption and produces a high covering
fraction out to the virial radius. There is more strong \HI\ from
recycled accretion than pristine accretion.

\item Accreting material is not unenriched. Recycled accretion
dominates the mass and metal budget of accreting material, and this
component is significantly enriched.

\item~\ion{Mg}{ii} provides a good tracer of accreting material
that was once in a wind (``recycled accretion").  Virtually all
strong \ion{Mg}{ii} absorbers arise in this category of gas.
\ion{Si}{iv}, despite its high ionisation level, behaves more like
\ion{Mg}{ii} than a high ion.

\item~\ion{C}{iv} represents an intermediate ion.  At small impact
parameters it mostly arises in recycled accretion, but at larger
impact parameters it mostly arises from ancient outflows.  The
crossover impact parameter is $\sim 100$~kpc in $L^*$ halos, and
moves to smaller impact parameters for smaller halos; for our lowest
mass halos, \ion{C}{iv} predominantly arises in ancient outflows
at all impact parameters.

\item~\ion{O}{vi} and \ion{Ne}{viii} almost exclusively trace ancient
outflows, at nearly all impact parameters and halo masses (very small impact
values around \mhi~halos being the exception).  A substantial
fraction of strong \ion{O}{vi} halo absorbers come from transition
gas at $\sim 300,000$~K and is cooling out of hot gas near
the virial temperature. Photoionised gas at $\sim 10^4$K also contributes
many \ion{O}{vi} absorbers, and most of the absorbers at impact parameters $\ge$ 100~kpc, in agreement with \cite{opp09}.

\item In small halos the majority of the accreting mass at $z=0.25$
is ejected back into a wind by $z=0$, with only about a quarter of
the mass remaining as stars or in the ISM.  In larger halos, by
$z=0$ the accreting mass is split evenly between stars, ISM gas,
and gas re-ejected as a wind.

\item With our definition of accretion (at $z=0.25$) as any gas
that will join a galaxy by $z=0$, velocity is not very useful for
distinguishing accreting gas from outflowing gas. The radial velocity
distribution of gas in the recycled accretion, young outflow, and
ancient outflow categories is very similar, with the exception that
material moving at high outward velocities at small radii does tend
to be recently ejected material.

\item Accreting material is not necessarily parallel to the disk,
and outflowing material is not necessarily perpendicular to the
disk.

\item Of gas not in the ISM at $z=0.25$, 86\% of the mass and 6\% of the metal mass has never been in a wind by $z=0$. 

\end{enumerate}

By decomposing our simulated gas distribution into inflows and outflows, we better understand
how the material in the CGM by $z=0.25$ arose, and what its eventual
fate is. These results underscore the dynamic nature of the CGM,
including how much material is participating in the ``baryon cycle",
particularly in halo fountains. This means that accreting material
is often enriched, and inflows can be traced using metal lines. We
emphasize that low ions trace fundamentally different gas (cold,
dense, accreting) than high ions (warm, diffuse, from ancient
outflows not accreting), and that inflows and outflows have different
distributions both in physical and temperature-density phase space.
While there is often co-location of gas in various phases giving
rise to both low and high ions, it is dangerous to assume that it
arises in a single-phase gas.
\HI\ has more strong absorption from recycled accretion
than pristine accretion, which simply reflects the fact that
recycled accretion is dominating the inflow mass budget at low
redshifts in our models.

There are clear observational diagnostics amongst ambient material,
inflows, and outflows, and even between young ($\le$ Gyr) and ancient
($>$ Gyr) outflows. In general \HI\ traces ambient material,
\ion{Mg}{ii} and \ion{Si}{iv} trace recycled accretion, \ion{C}{vi}
traces both recycled accretion and ancient outflows depending on
impact parameter, with \ion{O}{vi} and \ion{Ne}{viii} tracing ancient
outflows. We also find differences between low ($<$${10}^{11.5}$
$\msolar$) and high ($>$${10}^{11.5}$ $\msolar$) mass halos, because
high-mass halos suppress infall owing to their higher halo gas temperatures.
Assembling these diagnostics into a full census of CGM gas remains
a work in progress, but 
these results can aid in interpreting and understanding absorption-line
observations of the
CGM.

This paper is the second in a series \citep[][being the first]{for13}
to confront successful models for galaxy-IGM coevolution with
absorption line observations of CGM gas. With direct comparisons
to observations from COS soon forthcoming, and ongoing improvements
to our simulation methodologies, we hope to answer such questions
as ```what is the most realistic description for outflows in
simulations?", ``what is the phase space structure of the CGM, and
how patchy is it?'', ``how well mixed are the metals?'', ``what
fraction of outflows end up bound to the halo?" and ``how tightly
are the CGM and its host galaxy linked?". Through these types of
investigations, we aim to further constrain our simulations while
gaining additional insights into the CGM and the dominant physical
processes that impact it.

\section{Acknowledgements}

We thank Jason Tumlinson, Molly Peeples, Cameron Hummels, and
Annalisa Pillepich for useful discussion. Partial support for this
work came from NASA ATP grants NNX10AJ95G and NNX12AH86G, HST grants HST-GO-11598
and HST-GO-12248, NASA ADP grant NNX08AJ44G, NSF grants
AST-0847667, AST-0907998, AST-0908334, and AST-133514, and the
South African National Research Chairs program.  The simulations
used here were run on computing facilities owned by the Carnegie
Observatories. Computing resources used for this work were made
possible by a grant from the the Ahmanson foundation, and through
grant DMS-0619881 from the National Science Foundation.

\clearpage

\end{document}